\documentclass[traditabstract]{aa}  

\pdfoutput=1

\usepackage{graphicx}
\usepackage{txfonts}
\usepackage{epstopdf}
\usepackage{xspace} 
\usepackage{natbib}
\bibpunct{(}{)}{;}{a}{}{,}

\newcommand{\herschel}{\textit{Herschel}\xspace}

\newcommand{\kms}{\rm km\,s^{-1}}
\newcommand{\pc}{{\rm pc}}
\newcommand{\mx}{{\rm max}}

\begin{document}

\title{Slingshot Mechanism in Orion:
Kinematic Evidence For Ejection of Protostars by Filaments}

\author{Amelia M.\ Stutz  \inst{1}
  \and
  Andrew Gould  \inst{1,2}}

\institute{Max-Planck-Institute for Astronomy, K\"onigstuhl 17, 69117 Heidelberg, Germany \\
  \email{stutz@mpia.de}
  \and 
  Dept.\ of Astronomy, Ohio State University, 140 W.\ 18th Ave., Columbus, OH, USA\\
  \email{gould@astronomy.ohio-state.edu}
  }
\date{Received ; accepted }

\abstract{By comparing three constituents of Orion A (gas, protostars,
  and pre-main-sequence stars), both morphologically and
  kinematically, we derive the following conclusions. The gas surface
  density near the integral-shaped filament (ISF) is very well
  represented by a power law,
  $\Sigma(b) = 37\,M_\odot\,\pc^{-2}(b/\pc)^{-5/8}$ for the entire
  range to which we are sensitive, $0.05\,\pc < b<8.5\,\pc$, of
  projected separation from the filament ridge.  Essentially all Class
  0 and Class I protostars lie superposed on the ISF or on
  identifiable filament ridges further south, while almost all
  pre-main-sequence (Class II) stars do not.  Combined with the fact
  that protostars are moving $\la 1\,\kms$ relative to the filaments
  while stars are moving several times faster, this implies that
  protostellar accretion is terminated by a slingshot-like
  ``ejection'' from the filaments.  The ISF is the third in a series
  of identifiable ``star bursts'' that are progressively moving south,
  with separations of several Myr in time and 2--3 pc in space.  This,
  combined with the observed undulations in the filament (both spatial
  and velocity), suggest that repeated propagation of transverse waves
  through the filament is progressively digesting the material that
  formerly connected Orion A and B into stars in discrete episodes.
  We construct a simple, circularly symmetric gas density profile
  $\rho(r) = 17\,M_\odot\,\pc^{-3}(r/\pc)^{-13/8}$ consistent with the
  two-dimensional data.  The model implies that the observed magnetic
  fields in this region are subcritical on spatial scales of the
  observed undulations, suggesting that the transverse waves
  propagating through the filament are magnetically induced.  Because
  the magnetic fields are supercritical on scales of the filament as a
  whole (as traced by the power law), the system as a whole is
  relatively stable and long lived.  Protostellar ``ejection'' (i.e.,
  the ``slingshot'') occurs because the gas accelerates away from the
  protostars, not the other way around.  The model also implies that
  the ISF is kinematically young, which is consistent with several
  other lines of evidence.  In contrast to the ISF, the southern
  filament (L1641) has a broken power law, which matches the ISF
  profile for $2.5\,\pc < b<8.5\,\pc$, but is shallower closer in.
  L1641 is kinematically older than the ISF.  }

\authorrunning{A.\ Stutz \& A.\ Gould}  

\maketitle

\begin{figure}
\scalebox{0.57}{\includegraphics[trim = 0mm 0mm 0mm 0mm, clip]{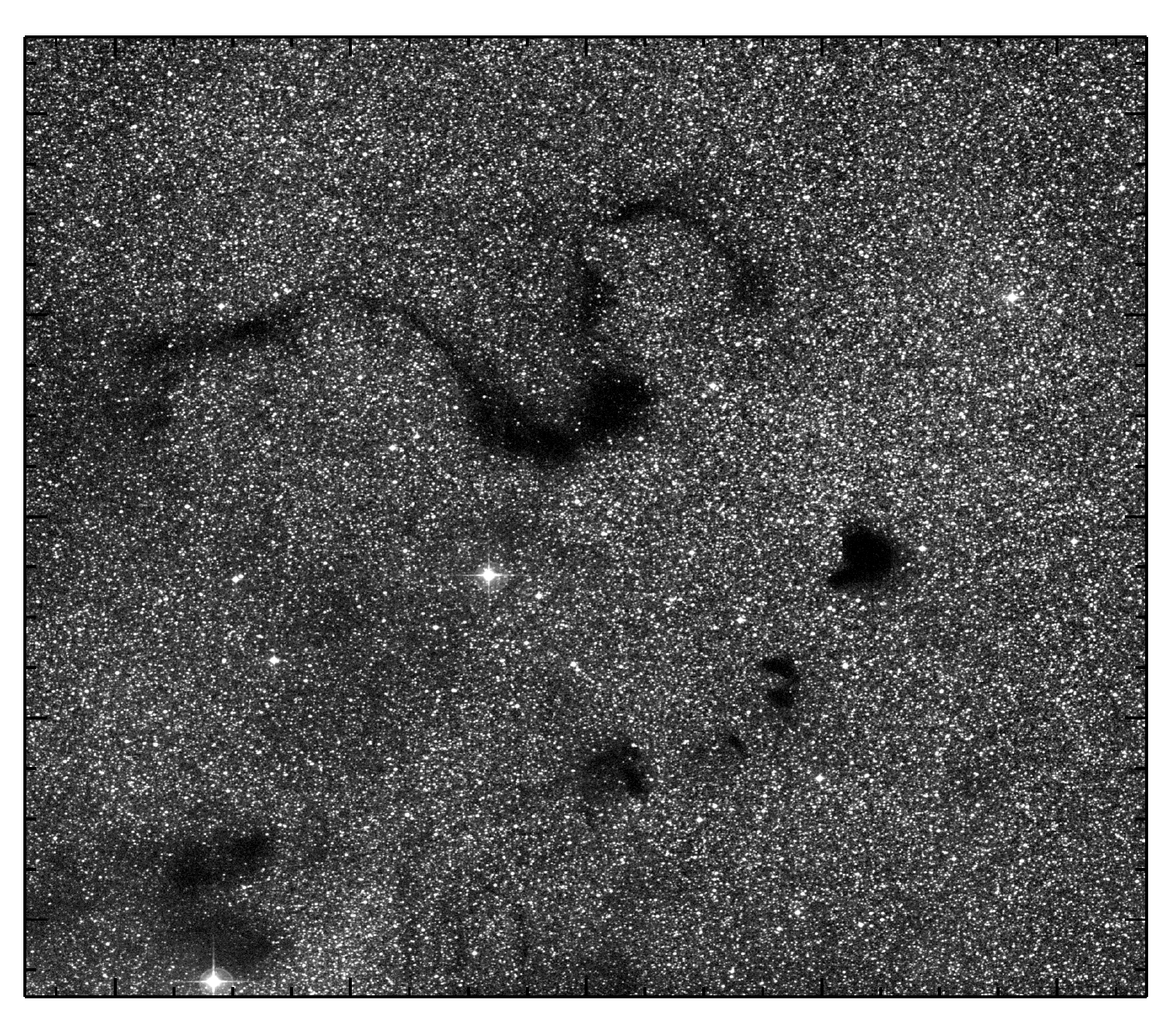}}
\begin{center}
\caption{Photographic image of Barnard 68 from Digitized Sky Survey.
  \citet{barnard1905} already argued that the ``uniform width'' and
  ``twistings and turnings of dark lanes'' including this one ``must
  have some meaning beyond mere chance''.
  ($57^\prime\times57^\prime$, center ($17^h24^m,-24^\circ$), north
  up, east left.)  }
\label{fig:barnard}
\end{center}
\end{figure}

\section{Introduction}

Opaque filamentary structures have been recognized in the interstellar
medium (ISM) for well over a century.  As already pointed out by
Barnard, the undulating form and uniform width of these structures cry
out for an explanation: ``These [structures] are especially striking
in $\alpha=19^{\rm h}23^{\rm m},\ \delta=+10^\circ 25^\prime$, where
they cover a space over $1^\circ$ wide and form a rather complicated
system of twistings and turnings of dark lanes....The strange thing
about all such lanes is that they always are of uniform width
throughout their ramifications.  This must have some meaning beyond
mere chance'' \citep{barnard1905}.  See Figure~\ref{fig:barnard}.

While Barnard himself does not appear to have ever offered such an
explanation, it seems difficult to conceive of any other than
near-critical magnetic fields, i.e., systems with comparable magnetic
and gravitational potential energy.  Of course, by now it is well
known from cosmological simulations (which are well matched by
observations) that gravity alone can produce filamentary structures.
However, these never have Barnard's ``twistings and turnings'' because
such morphologies would be unstable under the influence of gravity
alone. They require a competition to gravity from a restoring force.

While the general understanding that stars form from collapsing clouds
of gas goes back more than 300 years (Kant 1755), the fact that such
clouds are embedded in gas filaments has only become clear over the
last 50 years with the discovery, and gradually improving measurement,
of Class~0 and Class~I protostars directly superposed on filamentary
structures in Orion, Aquila, Taurus,
etc.\ \citep[e.g.,][]{andre14,stutz15}.

The discovery by \citet{heiles97} that the Orion A filament (the
largest nearby star-forming such structure) is enveloped in a helical
magnetic field greatly clarified the nature of these filaments.  The
observations (their Fig.~24) show magnetic field lines changing
direction as they cross the filament, from into to out of the plane of
the sky.  Since these are one-dimensional (1-D) projections of an
intrinsically 3-D field structure, they cannot be uniquely interpreted
by themselves.  However, given that circular (or more generally,
helical) fields, which would be generated primarily by currents moving
along the filaments, are the form that is necessary to confine
filaments of approximately uniform thickness, the \citet{heiles97}
observations constituted a ``smoking gun''.  Moreover, polarization
measurements by \citet{matthews2000} provide information in a second
dimension that confirms this picture.  That is, under the assumption
that this polarization arises from dust grains aligned by either
paramagnetic inclusions or radiative torques, the field lines pass
over the filament perpendicular to its axis.  See their Figure~1.
Subsequent observations confirm these results
\citep[][]{poidevin10,poidevin11}.  See also \citet{pillai15} for dust
polarization examples in more distant and massive clouds.

The potential benefits of space-based observations from {\it Spitzer}
and especially {\it Herschel} are two-fold.  First, deep observations
over a broader range of wavelengths permit studies that go much deeper
into protostellar envelopes and hence probe much earlier phases of
stellar birth (e.g.,
\citealt{stutz10,ragan12,pezzuto12,stutz13,furlan14,safron15}).
Second, the stability of these space-based observations allows one to
probe to much lower column densities, and thus to understand the
gravitational potential that serves as governor to the system in the
face of sustained (or repeated episodes of) star formation.

While the first aspect has been systematically exploited, the second
is mostly untouched.  Ground-based data are intrinsically ``biased''
against low-column-density regions in two distinct, if related, ways.
First, high and variable atmospheric foregrounds fundamentally limit
the precision of surface brightness measurements and so the minimum
surface brightness that can be measured.  More subtlety, the very
process of removing this background (chopping) acts as a high-pass
spatial filter and therefore emphasizes the spiny filamentary features
at the expense of the much larger low-density structures that govern
the overall gravitational potential.

However, because of the high interest of the spiny filaments
themselves\citep[e.g.,][]{andre10,arzoumanian11}, particularly because
these are the immediate sites of star formation, space-based
morphological studies have primarily emphasized probing the fine
detail of these structures using wavelet analysis or other high-pass
filters, which deliberately reproduces the principal characteristic of
ground-based data.

\begin{figure*}
\scalebox{0.66}{\includegraphics[angle=-90]{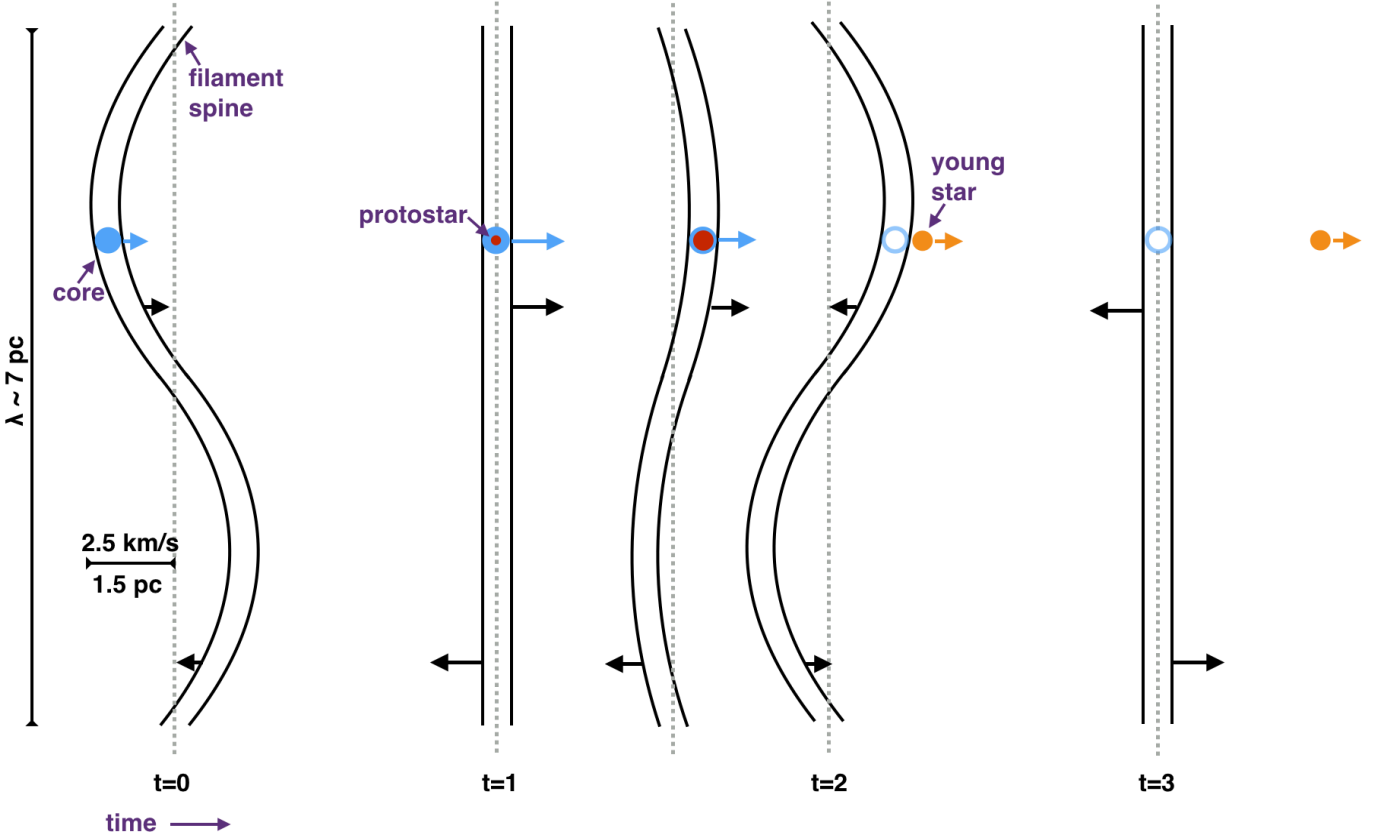}}
\begin{center}
  \caption{Cartoon showing time series of slingshot ``ejection'' of a
    protostar from the intergral shaped filament. Left to right: (1)
    core forms in filament; (2) core evolves into protostar and
    remains mechanically entrained in accelerating filament; (3)
    protostar continues to gain mass but remains in filament due to
    relatively low instantaneous filament acceleration; (4) filament
    reaches maximum acceleration (toward left) so that inertia of
    protostar causes it to leave filament, decoupling from dense gas
    cradle; (5) ``ejected'' young star moves away from filament.}
\label{fig:cartoon}
\end{center}
\end{figure*}

Here, we adopt an integrated, empirical approach to this subject.
We begin by focusing on the least studied aspect, the overall
morphology of the Orion~A cloud on the largest scales available
from {\it Herschel} data.  This allows us to estimate the gravitational
potential on all scales from the resolution limit (about 0.04~pc)
to 8.5~pc.  This sets the stage to examine the kinematics of the
stars and gas in a new light and to derive far-reaching conclusions.

In particular, we argue for a new ``slingshot mechanism'' that
``ejects'' protostars from the dense filaments that nurtured them,
thereby cutting off their accretion of new gas.  That is, the
filaments are always undergoing transverse acceleration, and the
nascent protostars are accelerated with them.  When the protostar
system becomes sufficiently massive to decouple from the filament, it
is released.  See Figure~\ref{fig:cartoon} for a schematic diagram of
the process.  As with a terrestrial hunter's slingshot, no impulse is
imparted to the projectile at the moment of release.  Rather, it is
the filament that accelerates away from the protostar.  As with the
hunter, so with the Hunter.

The reader with limited time is advised to skip directly to the
conclusions.

\begin{figure*}
\scalebox{0.66}{\includegraphics[angle =-90]{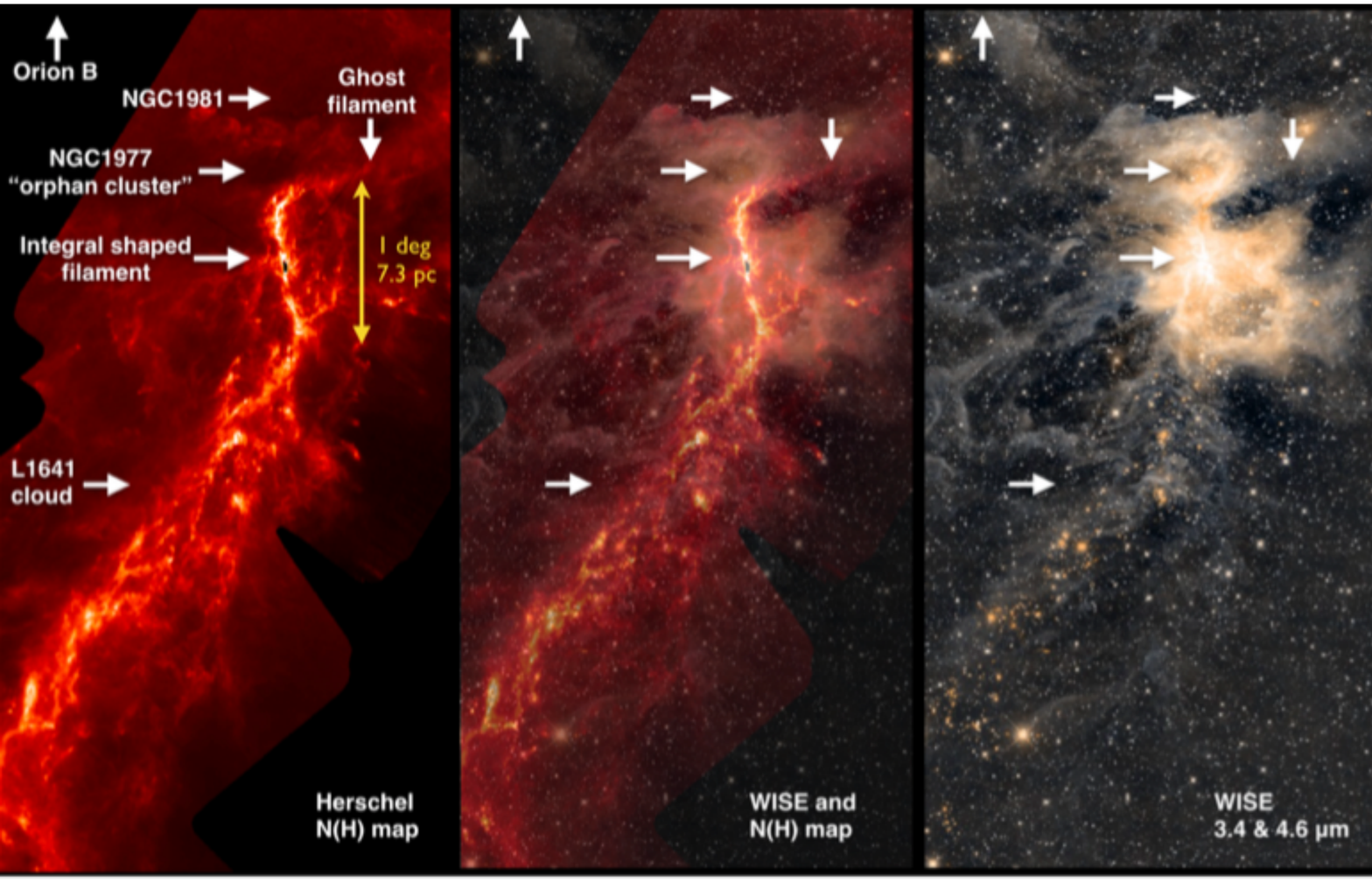}}
\caption{Location of the regions and structures discussed
  throughout this paper shown over the \herschel column density map
  (left), the WISE combined image \citep[][right]{lang14} and
  the combination of the two (middle).}
\label{fig:reg}
\end{figure*}

{\section{Power-Law Morphology of the Integral-Shaped Filament }
\label{sec:power}}

As its name implies, the integral-shaped filament (ISF) near the
northern end of Orion A \citep[e.g.,][]{bally87,johnstone99} is characterized by a
ridge of gas in the shape of an elongated ``S''.  See
Figure~\ref{fig:reg}.  For purposes of this
paper, we designate the ISF as $-5.9^\circ<\delta<-4.9^\circ$ and
refer to the remainder of Orion A ($-9.1^\circ<\delta<-4.9^\circ$) as
L1641.  In order to study the extended morphology of these structures,
we first identify a ``ridgeline'' of peak gas surface density, using
the dust column density map of \citet{stutz15} as a function of
Declination $\delta$, $\alpha_{\rm ridge}(\delta)$.  We obtain
comparable column density maps to those reported by
\citet{polychroni13} in L1641, \citet{ripple13} from CO measurements,
and \citet{lombardi14} using a combination of near-infrared extinction
and \herschel emission.  We then define $(x,y)$ positions by
\begin{equation}
x = (\alpha - \alpha_{\rm ridge})(\cos\delta) D_{\rm Orion},
\qquad
y = \delta\,D_{\rm Orion},
\label{eqn:xydf}
\end{equation}
where $D_{\rm Orion}=420\,$pc.

\begin{figure}
\scalebox{0.57}{\includegraphics{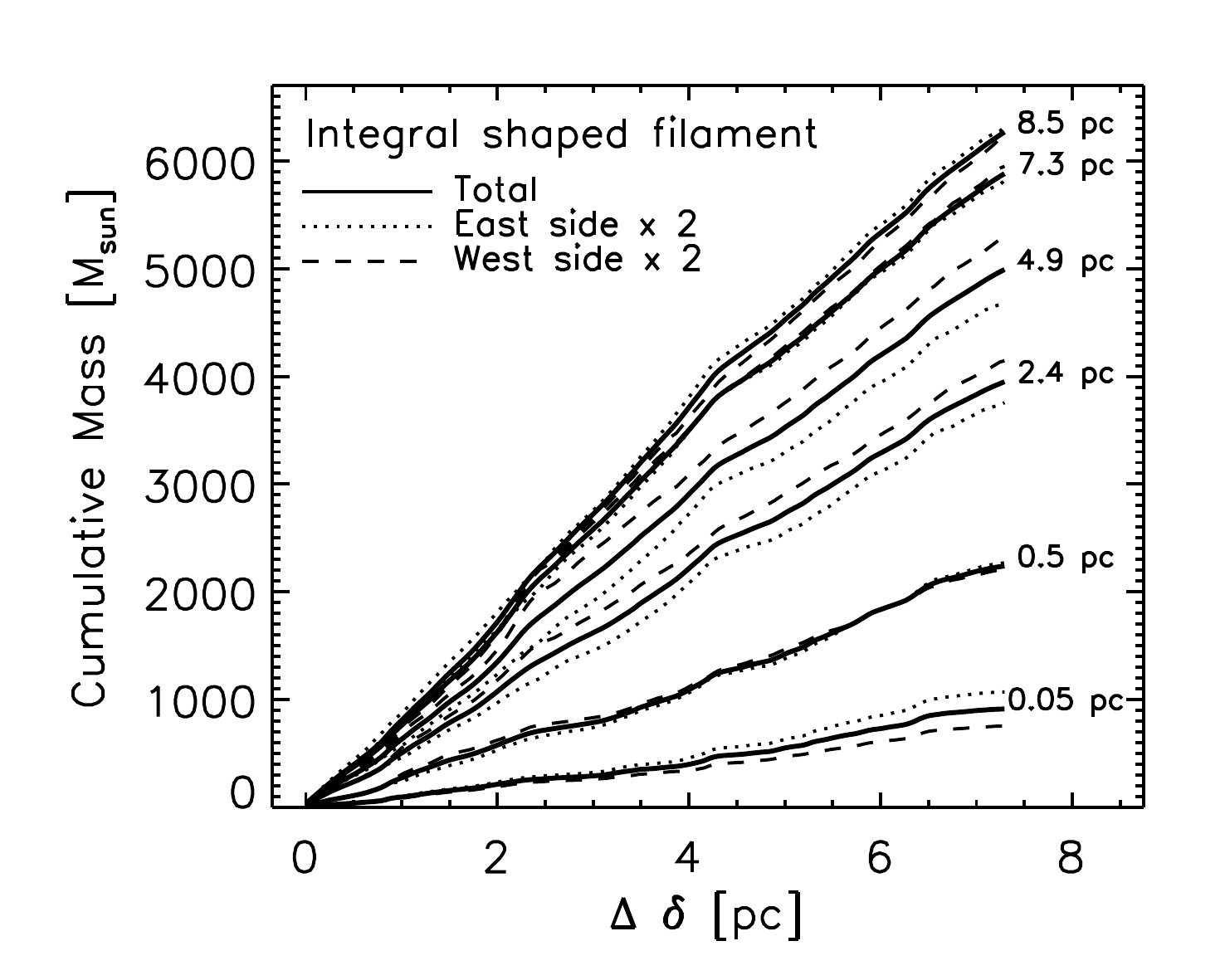}}
\scalebox{0.57}{\includegraphics{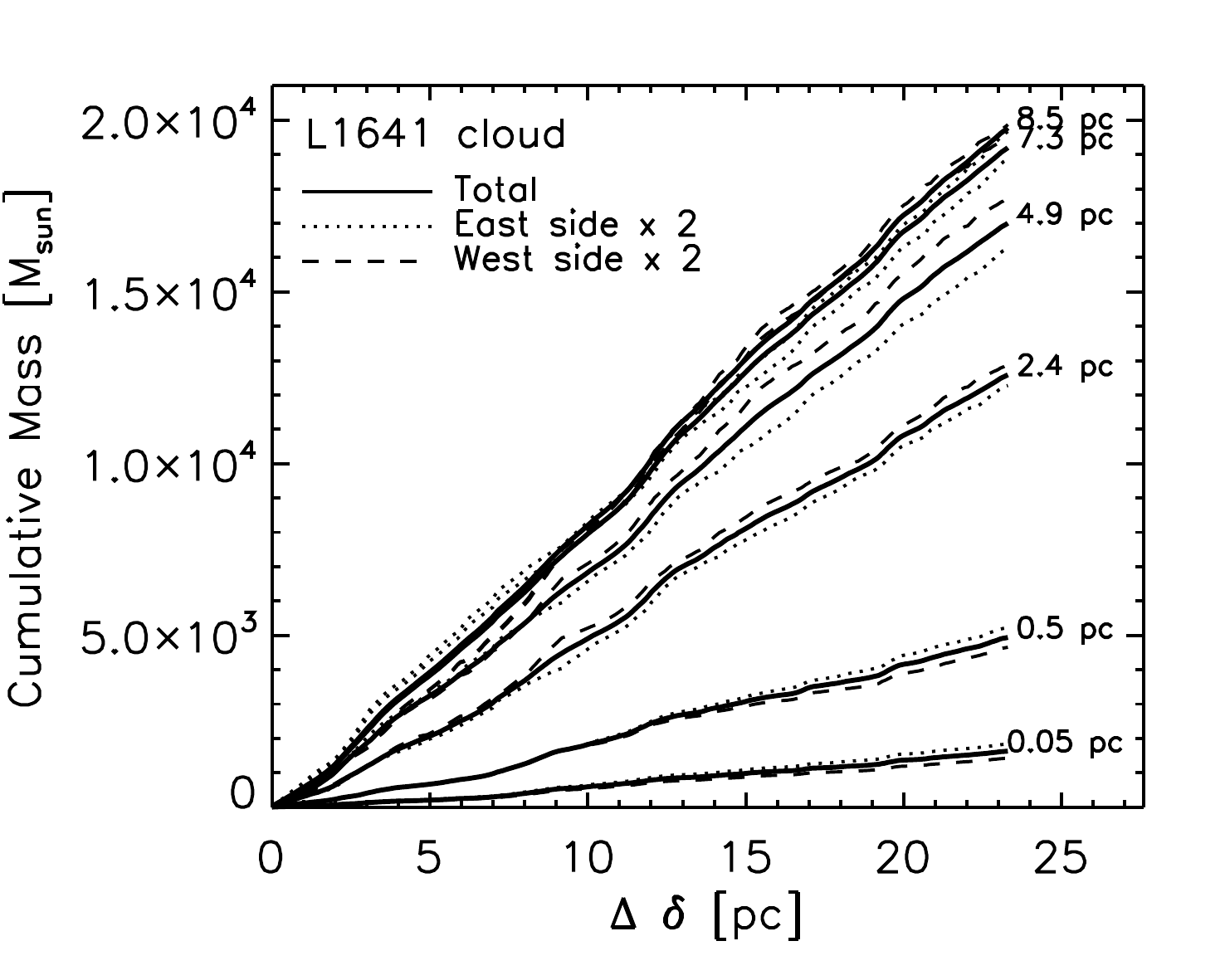}}
\caption{Cumulative distributions of mass within various projected
  separations $w$ from the density ridgeline.  Top and bottom panels
  show the Integral Shaped Filament (ISF) and remaining region of
  Orion A (L1641) respectively.  The cumulative distributions start at
  the southern boundary of each structure.  The separate cumulative
  distributions for the east $(-w<x<0)$ are west $(0<x<+w)$ are shown
  in different line types.  For ease of comparison (and clarity) these
  are multiplied by two.}
\label{fig:cum1}
\end{figure}

\begin{figure}
\scalebox{0.57}{\includegraphics{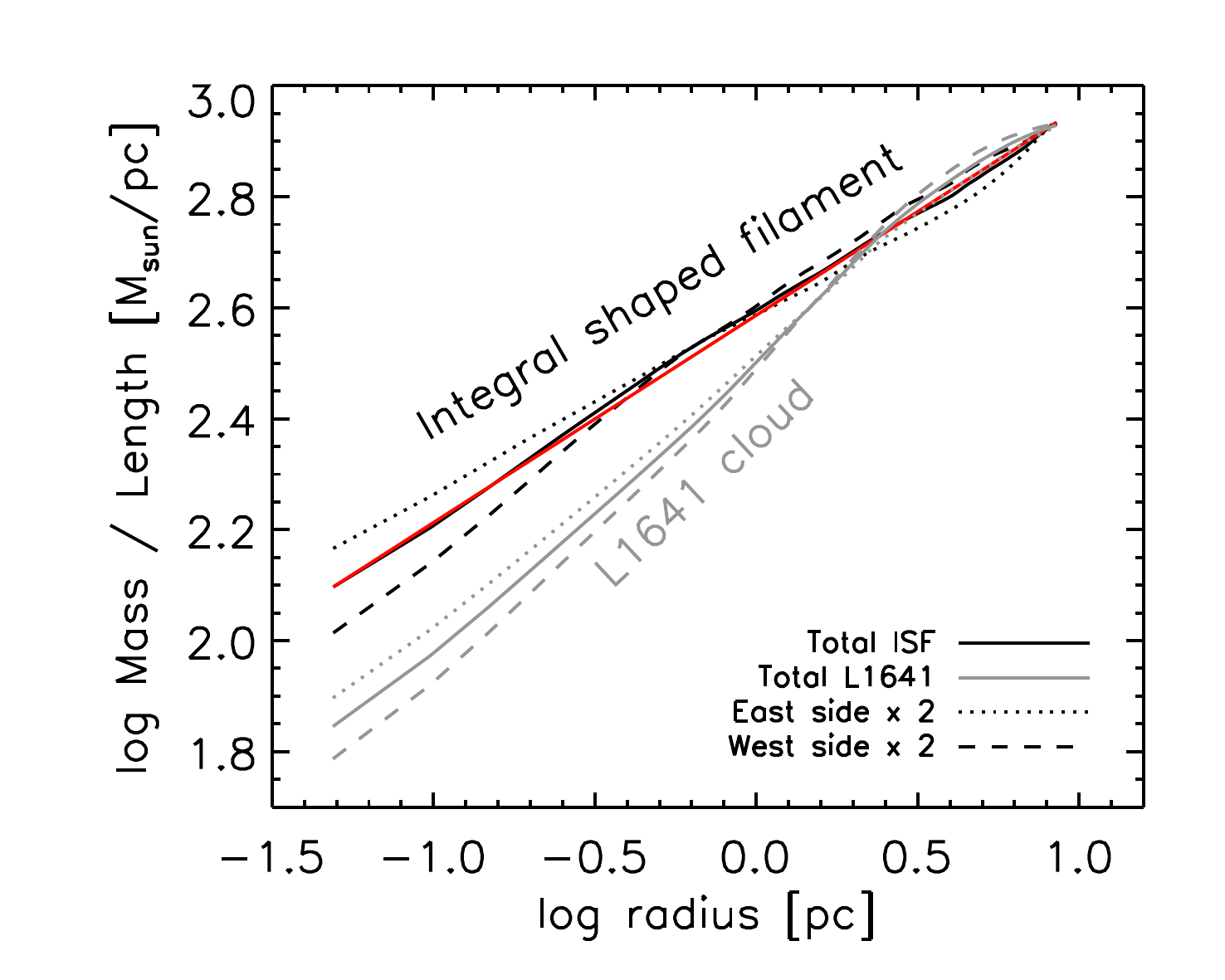}}
\caption{Enclosed projected mass per unit length as a function of
  enclosing width $w$, for the ISF and L1641, shown separately.  As in
  Figure~\ref{fig:cum1}, the east and west subsets are shown
  separately (after multiplying by 2).  The ISF is extremely well
  approximated by a pure power law, which is explicitly indicated as a
  red line.  L1641 shows the same power law (both slope and
  normalization) for $2.5\,\pc\lesssim w<8.5\,\pc$, but the surface
  density falls off substantially more slowly in the inner
  $2.5\,\pc$. }
\label{fig:cum2}
\end{figure}

We then form cumulative distribution functions of mass for various
different widths $w$ as a function of $\delta$
\begin{equation}
M(w,\delta) = 110\times 1.34\sum_{-y_0 < j < y}\sum_{-w<i<w} M_{ij,\rm dust},
\label{eqn:cum}
\end{equation}
where $y_0$ is the southernmost point of the structure in question,
$M_{ij,\rm dust}$ is the mass in dust of the $(i,j)$ pixel, 110 is the
assumed hydrogen-to-dust mass ratio, and 1.34 is the assumed ratio of
total gas to hydrogen gas. Figure~\ref{fig:cum1} shows this function
for several values of $w$.  These curves are all well approximated by
straight lines (as are the curves for all intervening values of $w$,
not shown).  This demonstrates that, at a given distance from the
ridgeline, the surface density is essentially independent of $\delta$.
This permits us to average over all $\delta$ within each structure to
find the line density $\lambda(w)$ as a function of enclosing width
\begin{equation}
\lambda(w) = {M(w,\delta_+) -M(w,\delta_-)\over D_{\rm Orion}(\delta_+ -\delta_-)} 
\label{eqn:lambdaA}
\end{equation}
where $\delta_\pm$ are the adopted boundaries of the structures,
indicated above.  For the ISF, this cumulative distribution function
is very well approximated by a straight line (on a log-log plot,
Figure~\ref{fig:cum2}), namely,
\begin{equation}
\lambda(w) = K\biggl({w\over\pc}\biggr)^\gamma;
\qquad K = 385\,{M_\odot\over\pc},
\qquad \gamma = {3\over 8}.
\label{eqn:lambda2}
\end{equation}

Note that this power-law behavior extends to $w=\pm 8.5\,\pc$, i.e.,
$\pm 1.2^\circ$, which is the approximate limit of the underlying
\citet{stutz15} map.  We note that the L1641 cumulative distribution
is a broken power law.  For $w>2.5\,\pc$, both the slope and
normalization are consistent with those of the ISF
(Equation~(\ref{eqn:lambda2})), while at closer separations the
projected density falloff is significantly more shallow $\gamma({\rm
  L1641},<2.5\,\pc)-1= -1/2$ versus $-5/8$ for the ISF.

{\section{Inferred Volume Density and Gravitational Potential}
\label{sec:grav}}

Inspired by the simplicity of Equation~(\ref{eqn:lambda2}), we assume
axial symmetry about the ridgeline.  Then, after some algebra, we find
local volume densities of\footnote{Almost all integrals in this paper
  are reducible to $\int_0^1 dx\, x^a (1-x)^b=a! b!/(a+b+1)!$}
\begin{equation}
\rho(r) = 
{\gamma(-\gamma/2)!\over 2(-\gamma/2-1/2)!(-1/2)!}\,{K\over\pc^2}
\biggl({r\over\pc}\biggr)^{\gamma-2},
\label{eqn:rho}
\end{equation}
and consequently enclosed line density,
\begin{equation}
\Lambda(r) = \int_0^r 2\pi r'\,dr'\,\rho(r') = 
f(\gamma)\lambda(r);
\label{eqn:lambda3}
\end{equation}
\begin{equation}
f(\gamma)\equiv{(-\gamma/2)!(-1/2)!
\over (-\gamma/2-1/2)!}\rightarrow 0.711,
\label{eqn:lambda3p}
\end{equation}
where the arrow indicates evaluation
according to the measured parameters.  
For locations close to the filament, the acceleration is
(by Gauss's Law) $a=2G\Lambda/r$, and hence the gravitational
potential is
\begin{equation}
\Phi(r) = \int_0^r dr'\,a(r') =
\eta(\gamma)G\lambda(r)
\rightarrow 6.3(\kms)^2\biggl({r\over\pc}\biggr)^\gamma
\label{eqn:phi}
\end{equation}
\begin{equation}
\eta(\gamma) \equiv {2(-\gamma/2)!(-1/2)!
\over \gamma(-\gamma/2-1/2)!}\rightarrow 3.79
\label{eqn:phip}
\end{equation}

\begin{figure}
\scalebox{0.48}{\includegraphics[trim = 31mm 0mm 30mm 0mm, clip]{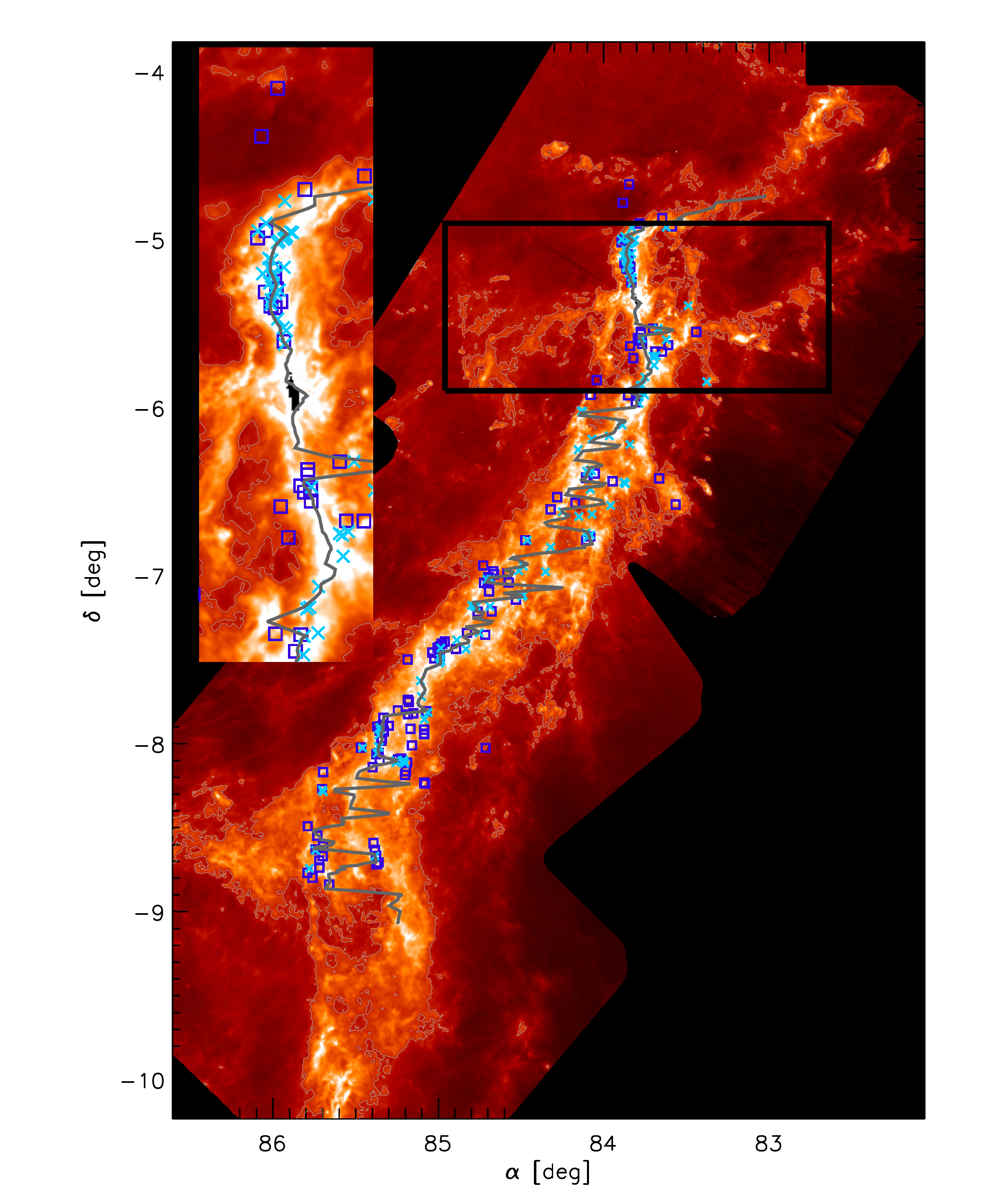}}
\caption{Positions of protostars \citep{ferlind15} compared to gas
  density as traced by dust emission \citep{stutz15}.  Protostars lie
  almost exclusively on the ridges of columns.  In the north, they are
  confined to the integral-shaped filament (ISF), which is the
  dominant structure.  In the south, the filamentary structure is more
  complex.  Ordinate is $\delta$ [deg].  Protostars with (without)
  radial velocity measurements are shown in cyan (blue). The
  approximate region used to calculate the potential is indicated with
  a black box of size $\sim$19$\times$7.3~pc$^{2}$.}
\label{fig:morph-proto}
\end{figure}

\begin{figure}
\scalebox{0.48}{\includegraphics[trim = 31mm 0mm 30mm 0mm, clip]{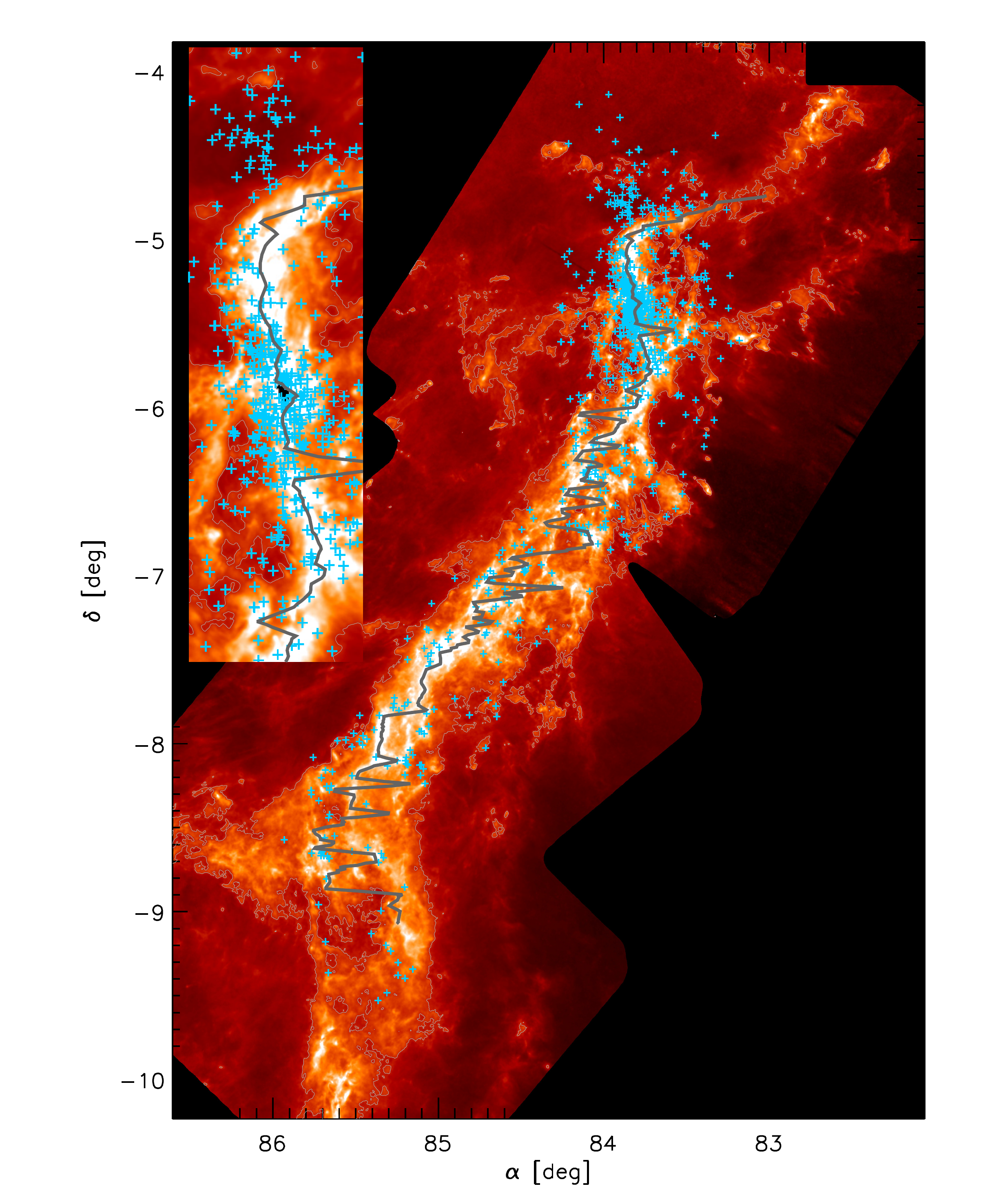}}
\caption{Positions of pre-main-sequence Class II stars (hereafter
  ``stars'') \citep{megeath12} compared to gas density as traced by
  dust emission \citep{stutz15}.  Only stars with radial velocities
  from Apogee are shown.  In contrast to the protostars
  (Fig.~\ref{fig:morph-proto}) the stars have essentially all left the
  filament (but see Fig.~\ref{fig:meg-non-apog}).  This dichotomy
  suggests that protostar accretion of ambient gas is terminated by
  ejection from the gas columns.  }
\label{fig:morph-star}
\end{figure}

We note that while the local density and the gravitational
acceleration diverge as $r\rightarrow 0$, the potential is well
behaved.  This means in particular that while the density profile must
flatten at some point, kinematics outside of the innermost pixel that
is measured (at about 10,000 AU) are insensitive to these details.

Next we calculate the time required for a star, initially at rest
at radius $r_\mx$ to fall to zero,
\begin{equation}
\Delta t = \int_0^{r_\mx} dt(r) = g(\gamma)
 {r_\mx\over \sqrt{2\Phi(r_\mx)}}
\label{eqn:deltat2}
\end{equation}
where
\begin{equation}
g(\gamma)={(1/\gamma)!(-1/2)!\over (1/\gamma-1/2)!}\rightarrow 3.03,
\label{eqn:ggamma}
\end{equation}
and
\begin{equation}
dt = {dr\over v(r)}, 
\label{eqn:dt}
\end{equation}
where
\begin{equation}
v(r) = \sqrt{2(\Phi(r_\mx)-\Phi(r))}, 
\label{eqn:vr}
\end{equation}
is the speed of a particle falling from $r_\mx$ to $r$.
Then, the mean value of any function $F(r)$ over the orbit is
\begin{equation}
\langle F\rangle = {\int_0^{r_\mx} dt(r)F(r)\over\Delta t}.
\label{eqn:vevf}
\end{equation}
In particular, the mean speed is
\begin{equation}
\langle v\rangle = {r_\mx\over\Delta t} = 
{\sqrt{2\Phi(r_\mx)}\over g(\lambda)}
\rightarrow 1.2\,\kms\biggl({r_\mx\over pc}\biggr)^{3/16},
\label{eqn:barv}
\end{equation}
while the mean radius is
\begin{equation}
\langle r\rangle = {r_\mx\over2}{(2/\gamma)!\over (2/\gamma-1/2)!}\,
{(1/\gamma-1/2)!\over (1/\gamma)!}\rightarrow {r_\mx\over\sqrt{2}}
\label{eqn:barr2}
\end{equation}
where the last step is strictly valid only in the limit of $\gamma\ll 1$,
but in practice is accurate to 2\% at $\gamma=3/8$.
For completeness, we note that the $n$th velocity moment is
given by
\begin{equation}
{\langle v^n\rangle\over [\Phi(r_\mx)]^{n/2}} = 
2^{n/2}{(n/2-1/2)!(1/\gamma - 1/2)!\over (-1/2)!(1/\gamma +n/2 - 1/2)!},
\label{eqn:barv2}
\end{equation}
so that in particular, for $n=2$, the right-hand side is
$2\gamma/(2+\gamma)\rightarrow 0.32$.

{\section{Orion A: Morphology and Kinematics}
\label{sec:m+k}}

{\subsection{Morphology: Gas vs.\ Protostars vs.\ Stars}
\label{sec:morphology}}

Figures~\ref{fig:morph-proto} and \ref{fig:morph-star} show the gas
column density as traced by dust emission in Orion A \citep{stutz15}
with, respectively, Class 0 and Class I protostars \citep{ferlind15}
and pre-main-sequence Class II stars \citep{megeath12} superposed.
The pre-main-sequence stars (hereafter ``stars'') shown are restricted
to the subsample with Apogee \citep{apogee} radial velocities for ease
of comparison with the figures in Section~\ref{sec:kinematics}.  All
the protostars are shown, but those without radial velocities are
displayed in a darker color.  The ridgeline is formed by finding the
$\alpha$ of the pixel with peak emission at each $\delta$ in the
4$\times$4 pixel (final pixel scale of $24^{\prime\prime}$) rebinned
dust map.

The key morphological features are
\begin{enumerate}
\item[1]{The protostars almost all sit on narrow gas-column ``ridges'',
while the stars are either more broadly distributed near the ridges,
or are displaced from the ridges altogether.}
\item[2]{The gas, protostars, and stars all have highest density along
integral-shaped filament (ISF, $\delta\sim -5.5^\circ$).}
\item[3]{There is a strong knot of stars, NGC 1977, just north of ISF,
  which is almost devoid of gas, and which we therefore dub the
  ``orphan cluster''.}
\item[4]{There is a tenuous gas ``ghost filament'' extending 
west and then northwest from northern tip of the ISF.}
\end{enumerate}

From the fact that the great majority of protostars, which are by
definition deeply embedded in dense gas, are superposed on narrow gas
column ridges, we can conclude that their envelopes are lost either
before or almost immediately after they lose access to the dense gas
column.  This directly raises the question of whether there is a
substantial population of stars within the gas column ridges.  If
protostellar envelopes are lost ``before'' leaving the filament then
there should be such a populations, while if they leave ``immediately
after'' then there should not.

Although it is not proved from Figure~\ref{fig:morph-star} alone,
there are in fact essentially no stars in Figure~\ref{fig:morph-star}
that sit directly on the gas column ridges.  That is, particularly for
the ISF, there are a considerable number of stars close to the
filament, so that just by chance some will be projected directly on
the narrow filament spine (i.e., the location of most of the
protostars).  However, we will show below, using radial velocity data,
that these projections are indeed almost all by chance.  These two
facts (one demonstrated and one anticipated) together argue that the
Class I (embedded) phase is ended by removal of the protostar from the
gas column ridge, rather than by somehow exhausting or cutting off
supply from the local gas column.

\begin{figure}
\scalebox{0.40}{\includegraphics[trim = 20mm 0mm 0mm 0mm, clip]{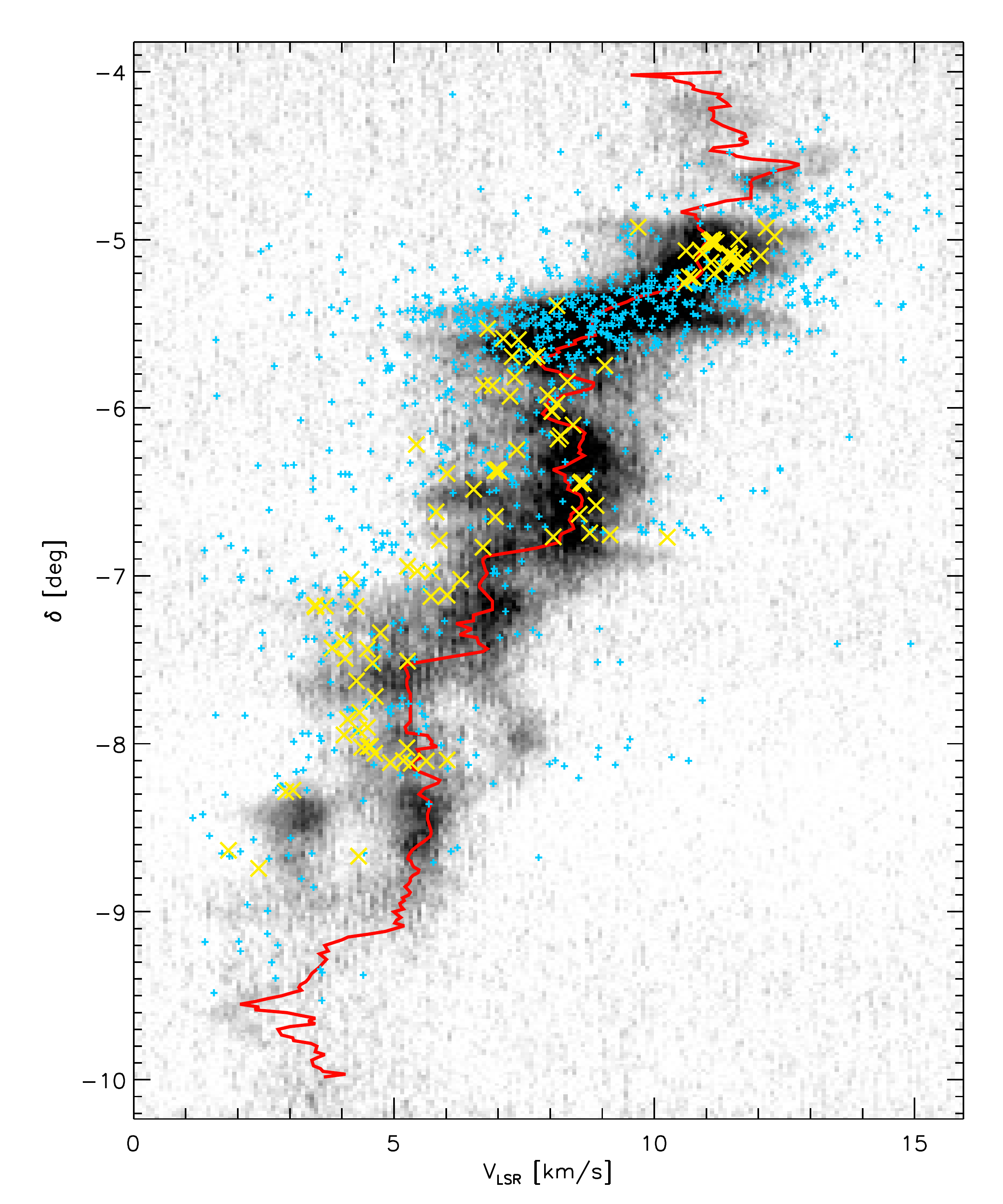}}
\caption{Position-velocity diagram of $\rm ^{13}CO$ gas (grey-scale),
  protostars (yellow $\times$-symbols), and stars (blue $+$-symbols).
  The red line indicates the $\rm ^{13}CO$ gas ridgeline
  (see~\ref{sec:kinematics}). Gas velocity profile is integrated over
  $\alpha$.  Protostars generally have velocities very close to the
  peak velocity of the local gas, whereas stars are further displaced.
  That is,  stars have both greater kinetic and greater potential energy
  than the protostars relative to their local gas columns.  Ordinate is
  $\delta$ [deg].  }
\label{fig:kinemat-all}
\end{figure}

Since these stars used to be protostars, they were formerly lying
directly on the gas-column ridge, just as the present-day protostars
are.  This would be inconsistent with Newton's First Law unless either
the stars or the gas column (or both) have accelerated between their
Class I phase and today.  If the gas were not accelerating, there
would be no way to accelerate the stars, since the line-density of
protostars and space density of stars are at least an order of
magnitude too low to induce transverse motions of the required
amplitude, particularly in the southern filament.  Hence, we conclude
that the filament is undergoing accelerated motion transverse to the
filament.  We will estimate the amplitude of this motion further
below.

The ``orphan cluster'' (NGC 1977) \citep{peterson08} lying just north
of the ISF must have formed out of filamentary gas, just like all the
other stars.  In principle, these could have been ejected
(accelerated) from the location of the present-day ISF.  If this is
the case, however, it will soon become apparent from Gaia data, since
the implied proper motions are of order $1\,\rm mas\,yr^{-1}$.  More
likely, the ISF extended to the current midpoint of the ``orphan
cluster'' a few Myr ago when these stars formed and has since been
dispersed or moved.  The extended ghost filament to the west is a good
candidate for the fate of this putative extension of the ISF.  In this
case the torques propagating northward through the filament
encountered suddenly weakening restoring forces as they exited the ISF
and simply ripped the filament away from its former trajectory toward
Orion B.


While the ``orphan cluster'' is heavily populated by ``disk sources''
(i.e., those found by \citealt{megeath12} to have {\it Spitzer}
infrared (IR) excesses) there is another cluster (NGC 1981), which
lies 3 pc further to the north (hence closer to Orion B), and which
was identified by \citet{megeath13} as containing relatively few
\citet{megeath12} disk sources but a much larger number of
photospheric (Class III) stars, meaning that it is older.

\begin{figure*}
  \scalebox{0.4}{\includegraphics[trim = 20mm 0mm 0mm 0mm, clip]{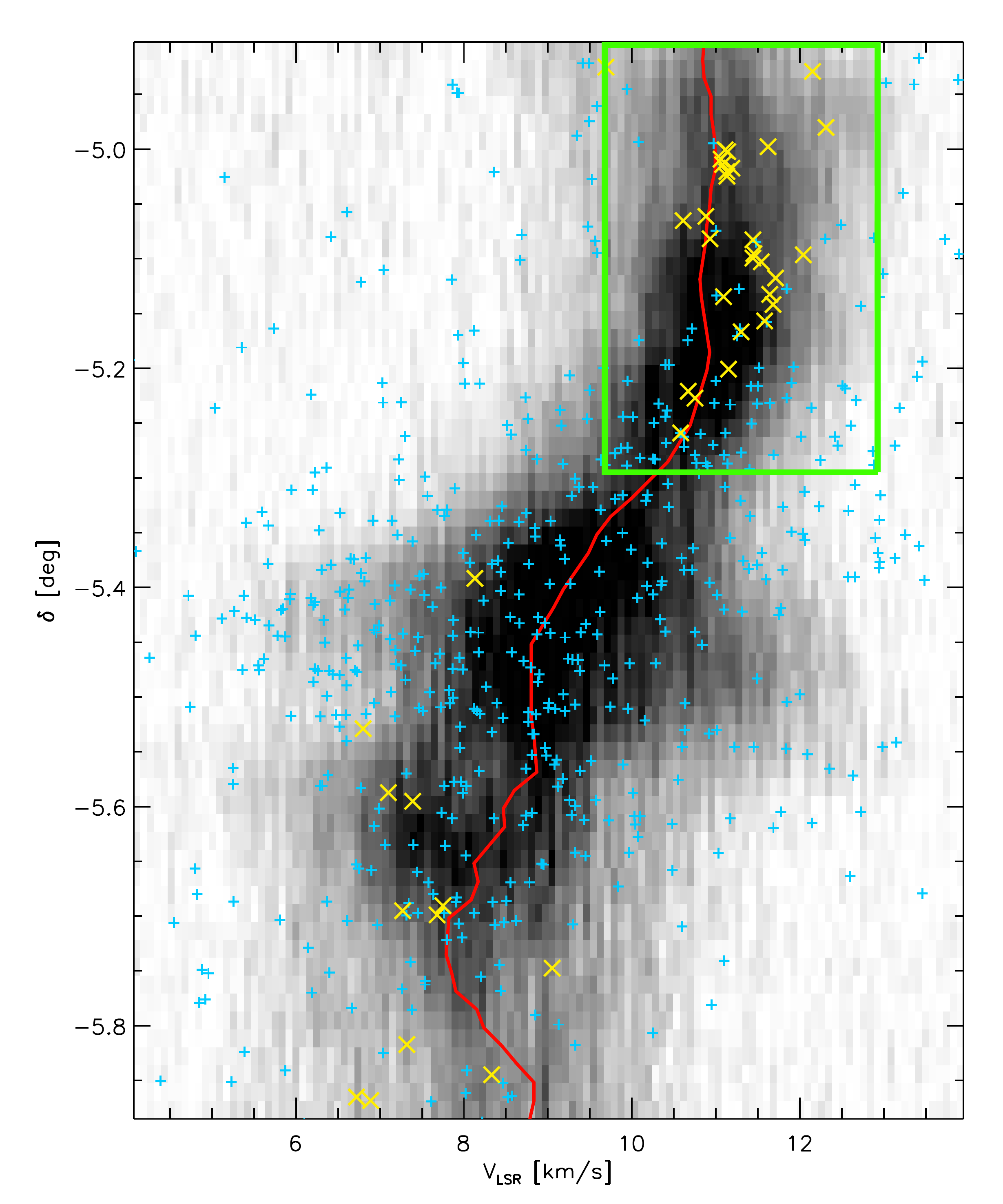}}\scalebox{0.4}{\includegraphics[trim = 20mm 0mm 0mm 0mm, clip]{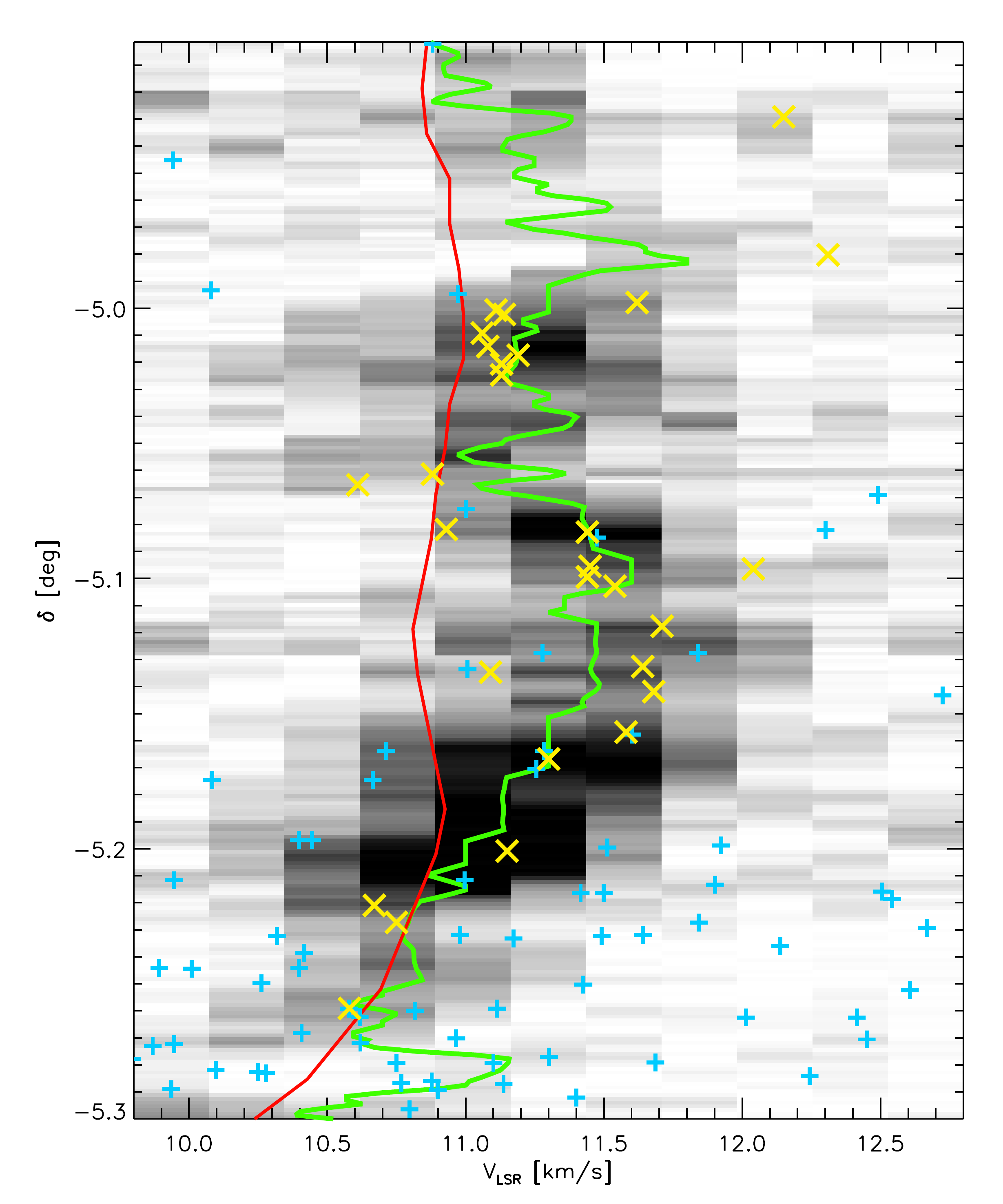}}
  \caption{Left: Zoom-in of position-velocity diagram shown in
    Figure~\ref{fig:kinemat-all}.  Green box indicates the area in the
    diagram on the right. Right: $\rm N_2 \rm H^+$ position-velocity
    diagram of the northern portion of the ISF.  Red line is the
    $\rm ^{13}CO$ gas ridgeline while the green line shows the {\it
      velocity centroid} of the $\rm N_2 \rm H^+$ line.  The
    protostars (yellow-$\times$ symbols) have velocities that are
    slightly displaced from the CO gas but directly associated with
    $\rm N_2 \rm H^+$.  Ordinate is $\delta$ [deg].}
\label{fig:kinemat-z}
\end{figure*}

Hence, the overall picture is of repeated episodes of star-cluster
formation, each triggered by a transverse wave that propagates
northward through the filament.  Each episode ends in a violent
contraction of the filament at its northern end (similar to the ISF
today), which both ignites an episode of star formation and then
disperses the ``loose end'' of the filament that formerly connected
Orion A and Orion B.

{\subsection{Kinematics: Gas vs.\ Protostars vs.\ Stars}
\label{sec:kinematics}}

Figure~\ref{fig:kinemat-all} is a position-velocity diagram for the
stars and gas shown in Figure~\ref{fig:morph-star} and for the subset
of the protostars shown in Figure~\ref{fig:morph-proto} for which
radial velocity (RV) data are available.  The stellar RVs are taken
from the Apogee catalog \citep{apogee,dario15}. See also
\citet{hacar16}. As mentioned in Section~\ref{sec:morphology} these
are restricted to stars having disks as classified by
\citet{megeath12}.

The initial sample has 1030 stars.  Roughly 1/4 of these have multiple
measurements and of these 15 show scatter excess of $2\,\kms$ and are
removed as binaries.  Because the RV distribution of this sample is
extremely compact, the 49 (out of the remaining 1015) stars with RVs
larger than $7\,\kms$ from the median are removed from the sample (as
being probable background contaminants, but perhaps binaries).  Note
that the Apogee sample is restricted to $H<12.5$, which may create a
subtle but important bias to which we return below.

The protostar RVs come from an online catalog \citep{difra16} of HOPS
sources, which contain independent measurements derived from
$\rm NH_3$ and $\rm HC_5N$.  If these agree to within $0.5\,\kms$ then
the $\rm N H_3$ measurement is accepted.  Only one star is eliminated
by this check.  Of the 248 protostars in Figure~\ref{fig:morph-proto},
111 have RV measurements.  The $\rm NH_3$ velocities agree well with
independent $\rm N_2 \rm H^+$ measurements
\citep{tatematsu08,tatematsu16} in the ISF, see
Figure~\ref{fig:kinemat-z}.

The gas velocities are derived from $\rm ^{13}CO (2-1)$
\citep{nishimura15} by collapsing the 3-dimensional (3-D) data cube in
($\alpha,\delta,v_r$) along the $\alpha$ axis.  The ridgeline is
formed by finding the velocity of the pixel with peak emission at each
$\delta$ and then boxcar smoothing with a length of 4 pixels
corresponding to $4^\prime$.  The $\rm ^{13}CO$ velocities and
ridgeline morphology are compared to the $\rm N_2 \rm H^+$
measurements in Figure~\ref{fig:kinemat-z}.  In the right panel we show
the CO gas ridgeline compared to the $\rm N_2 \rm H^+$ velocity
centroid calculated within 0.5 $\kms$ the isolated hyperfine component
line peak.  The protostars are substantially closer to the
$\rm N_2 \rm H^+$ velocity centroid than the CO gas ridgeline.  The
$\rm N_2 \rm H^+$ map coverage and sensitivity prevents detailed
analysis of larger areas of the ISF or L1641.

The main features of Figures~\ref{fig:kinemat-all} and \ref{fig:kinemat-z} are
\begin{enumerate}
\item[1]{The protostars are displaced from the gas column in RV by
  $\la 1\,\kms$.}
\item[2]{The stars are also generally displaced from the gas column
  but by several $\kms$.}
\item[3]{The gas column exhibits undulations in RV, particularly near
  the ISF, but also near $\delta\sim -6.5^\circ$, i.e., qualitatively
  similar to the morphological undulations.}
\item[4]{The orphan cluster (north of the ISF) is dispersed similarly
  to other stars.}
\end{enumerate}

\begin{figure*}
\scalebox{0.57}{\includegraphics[trim = 10mm 0mm 0mm 0mm, clip]{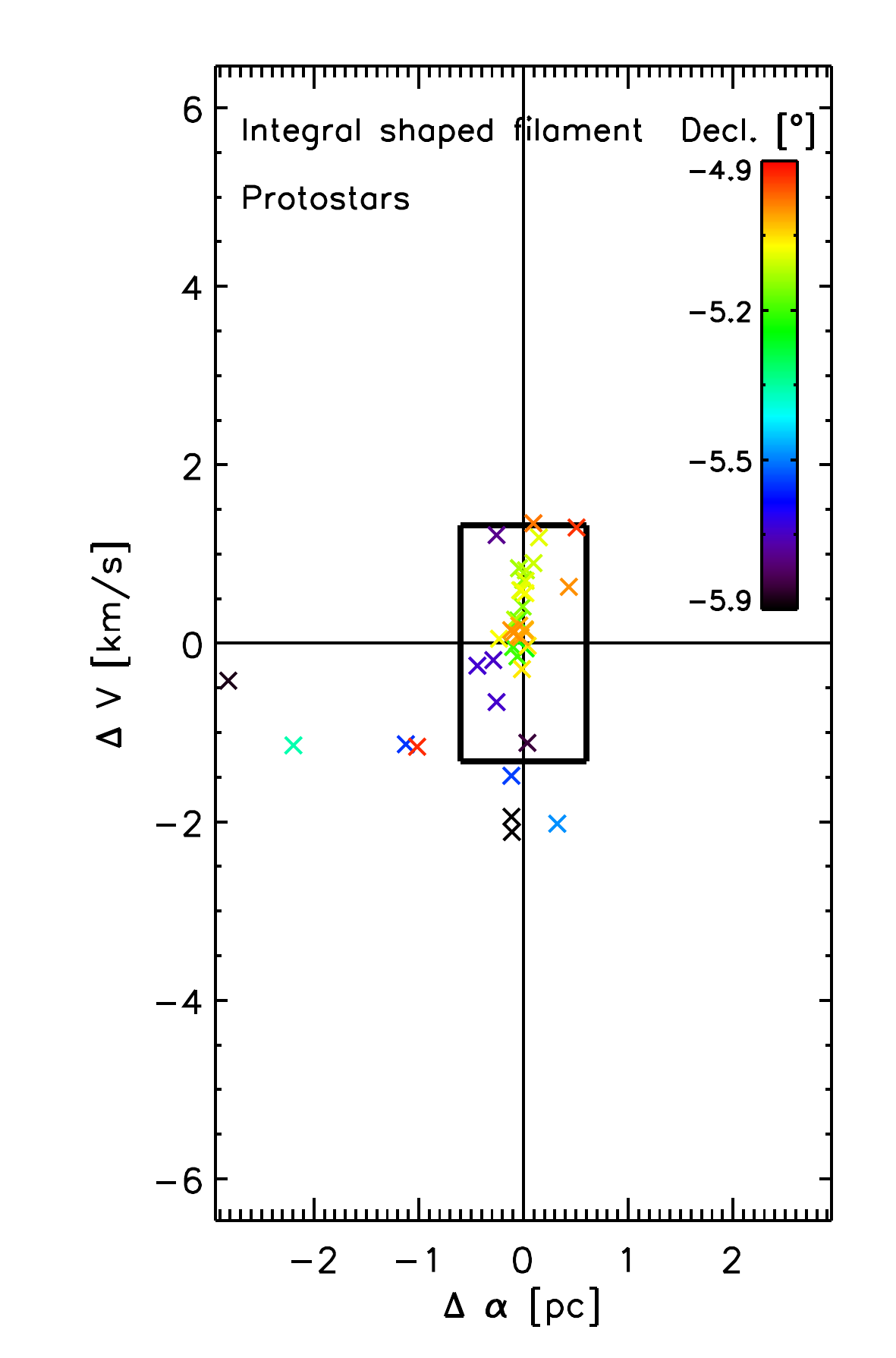}}\scalebox{0.57}{\includegraphics[trim = 18mm 0mm 0mm 0mm, clip]{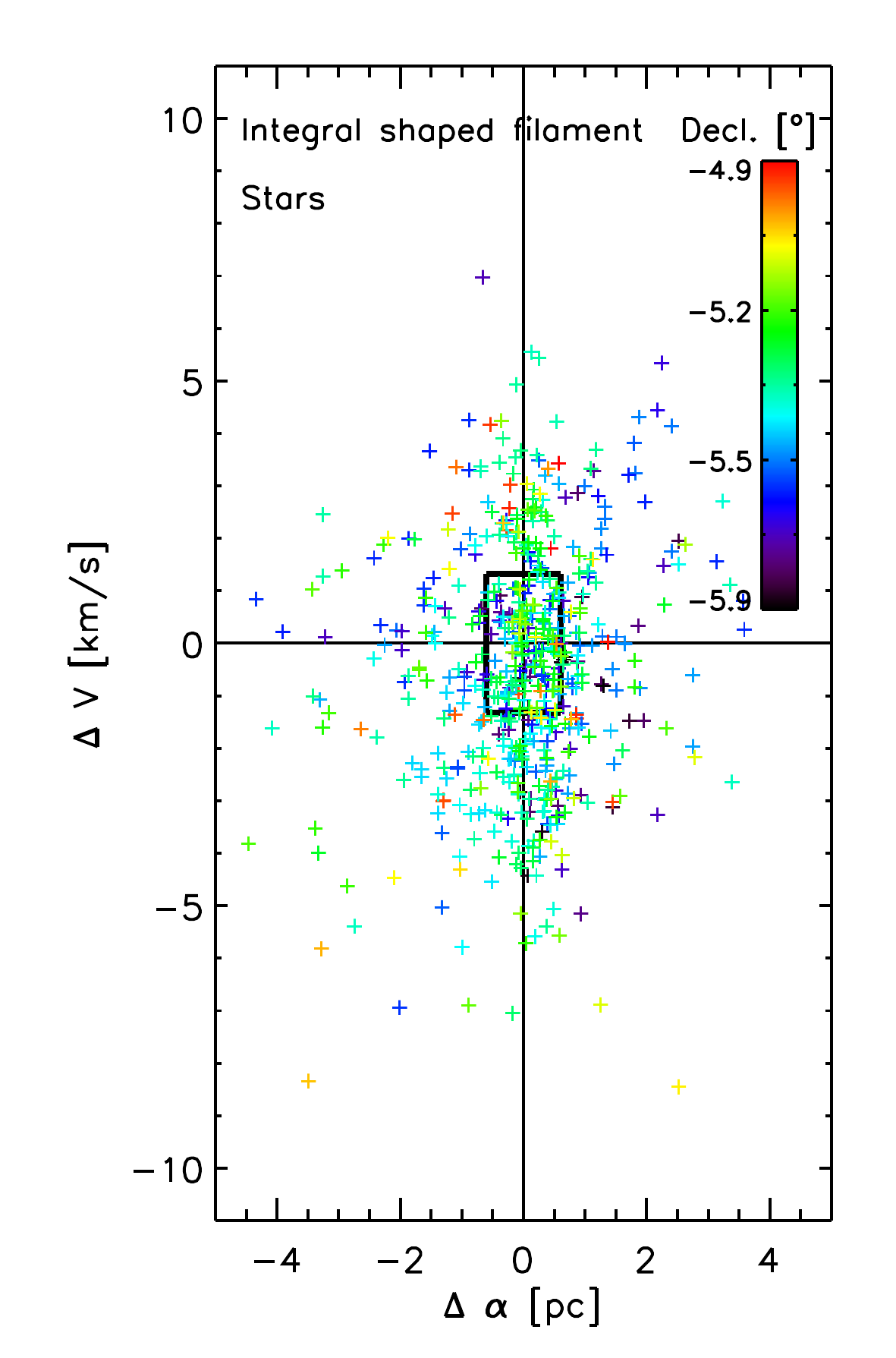}}\scalebox{0.57}{\includegraphics[trim = 18mm 0mm 0mm 0mm, clip]{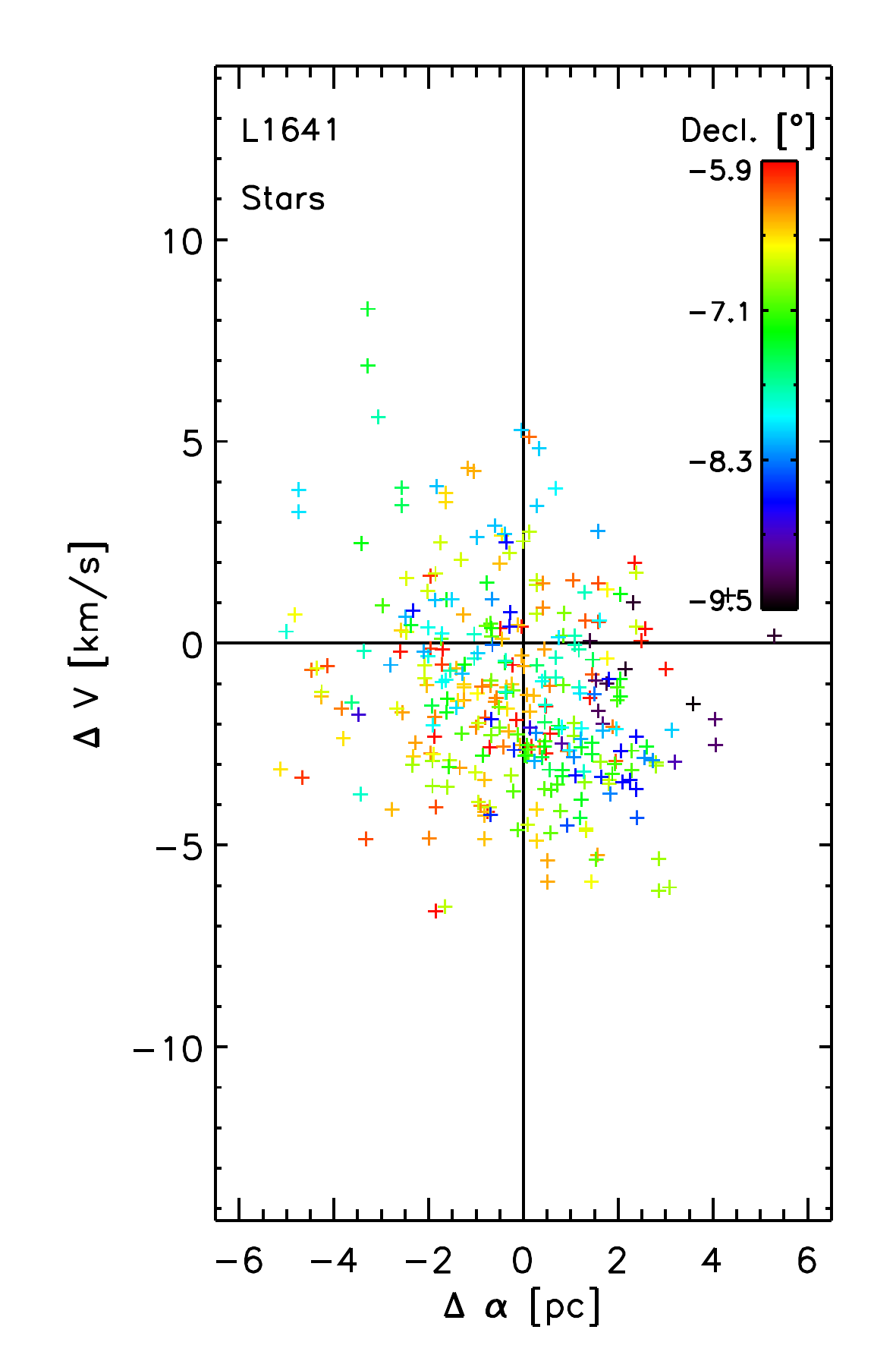}}
\caption{Offsets of ISF protostars (left), ISF stars (center), and
  L1641 stars (right) from the gas ridgelines in $\alpha$
  (Figs.~\ref{fig:morph-proto} and \ref{fig:morph-star}) and RV
  (Fig.~\ref{fig:kinemat-all}).  The protostars are shown only for the
  ISF because of the difficulty of capturing the filament structure
  further south by simple prescriptions. Points are color-coded
  north-to-south $\leftrightarrow$ red-to-blue. Ratio of axis scales
  is 1 Myr.  Hence, the vertically-elongated structure of these
  diagrams indicates timescales $\ll 1\,$Myr, $<1\,$Myr, and $\sim
  1\,$Myr for the three diagrams, respectively.  These could be either
  lifetime or turnaround time.  The stellar distributions (both ISF
  and L1641) are broadly distributed, with no obvious
  ``filament-hugging'' subpopulations, like the ISF protostars.
  However, they also show significant lumpiness demonstrating
  incomplete relaxation.  The protostars show some severe clumping,
  with a high density near the origin, in particular.  }
\label{fig:dradrv}
\end{figure*}

While it is clear that the stars are overall more displaced from the
filament than the protostars in both position and velocity, it is not
yet proved that there is no sub-population of stars that mimic the
protostars in being close to the filament in both these phase-space
coordinates.  To begin addressing this question we show in
Figure~\ref{fig:dradrv} the offsets in velocity versus offsets in
position for three populations, protostars in the ISF, stars in the
ISF, and stars in L1641.  We reserve discussion of most of the
features of these figures for Section~\ref{sec:charspeed}.  Here we
just point out that the ISF stars form a broad 2-D distribution in
this diagram, especially compared to the ISF protostars, similar to
the distributions observed by \citet{foster15} in the NGC 1333 region
of Perseus.  This confirms that there is no significant subpopulation
among the sample of \citet{megeath12} IR-excess (Class II) stars with
Apogee \citep{apogee} radial velocities that ``hugs'' the filaments in
the manner displayed by the protostars.

The fact that the protostars have substantially lower specific kinetic
energy (in addition to having lower potential energy, as indicated in
Section~\ref{sec:morphology}) confirms that nascent stars receive a
substantial kick between their protostellar and stellar phases.  As
discussed above, a natural explanation for this tight correlation
between evolutionary phase and total specific energy is that these
kicks remove the protostars from the gas rich column in which they
were born, thereby shutting off accretion.  However, one should also
consider the alternative explanation that gas accretion shuts down by
some self-generated process and the protostar-turned-star leaves the
column by some as yet unspecified process at a later time.  This
alternative appears to be strongly contradicted by the absence of
stars superposed on the ridges of gas columns.

\begin{figure}
\scalebox{0.50}{\includegraphics[trim = 31mm 0mm 40mm 0mm, clip]{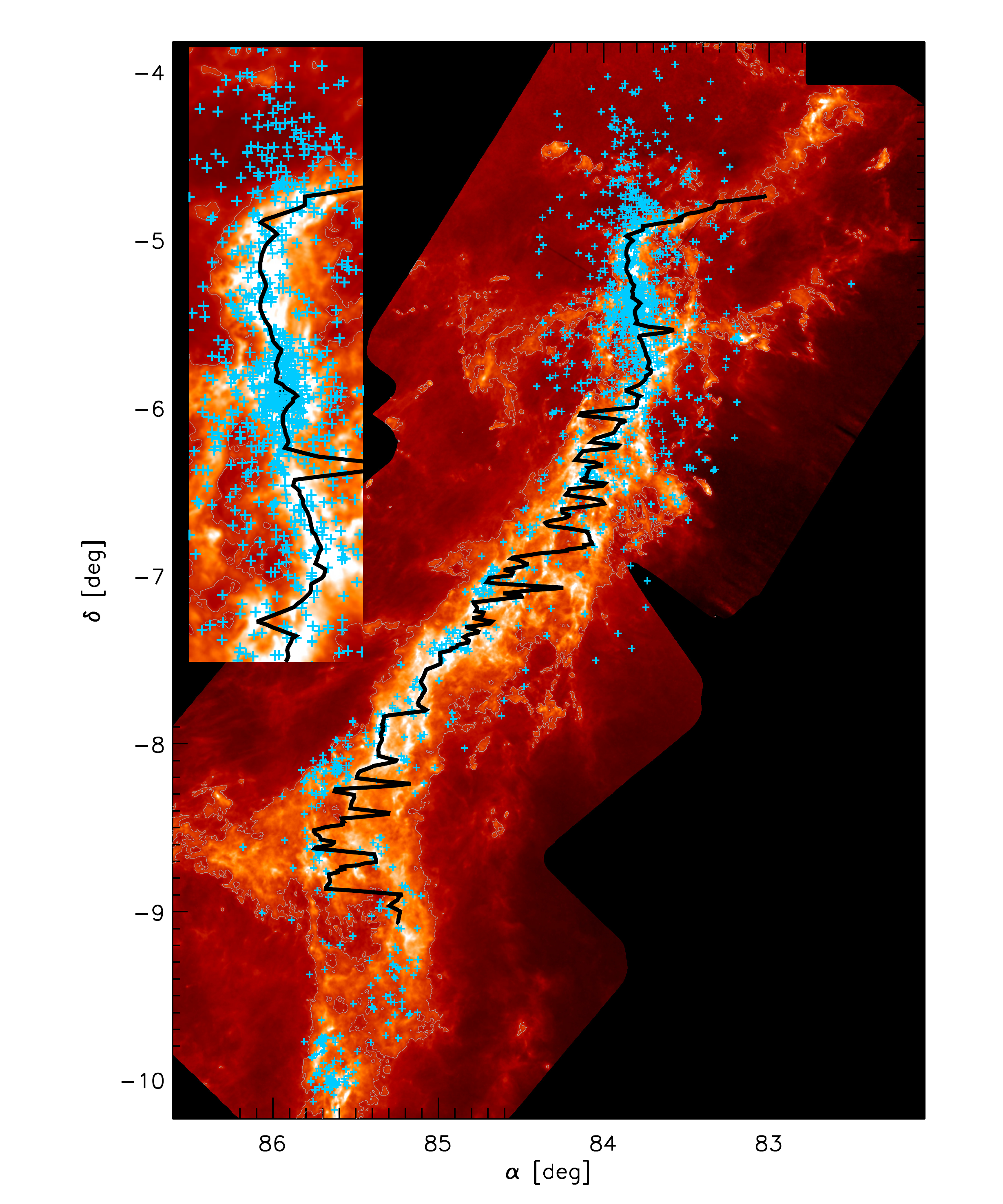}}
\caption{Star positions relative to gas for stars {\it without} Apogee
RVs, i.e., the stars that are excluded from Figure~\ref{fig:morph-star}.
We study these non-Apogee stars to determine whether the Apogee 
\citep{apogee,dario15} sample is systematically biased against
stars lying within the dense spiny columns due to increased
extinction.  See Figure~\ref{fig:non-apo-hist}.
Ordinate is $\delta$ [deg].
}
\label{fig:meg-non-apog}
\end{figure}

However, this seeming evidence could in principle be partly the result
of selection.  Class II stars that remained embedded in dusty columns
would be fainter in $H$ band due to extinction and so much less likely
to be observed by Apogee, given its $H<12.5$ limit.  In
Figure~\ref{fig:meg-non-apog}, we therefore show the \citet{megeath12}
stars that were {\it not} observed by Apogee.  For the most part,
these excluded stars are not lying on the spines of filaments, which
reflects the fact that there are heterogeneous reasons why
\citet{megeath12} stars were not observed by Apogee. However, there
may be some ``preference'' for stars very close to filaments, relative
to Figure~\ref{fig:morph-star}.  

\begin{figure*}
\scalebox{0.70}{\includegraphics[trim = 10mm 0mm 0mm 0mm, clip]{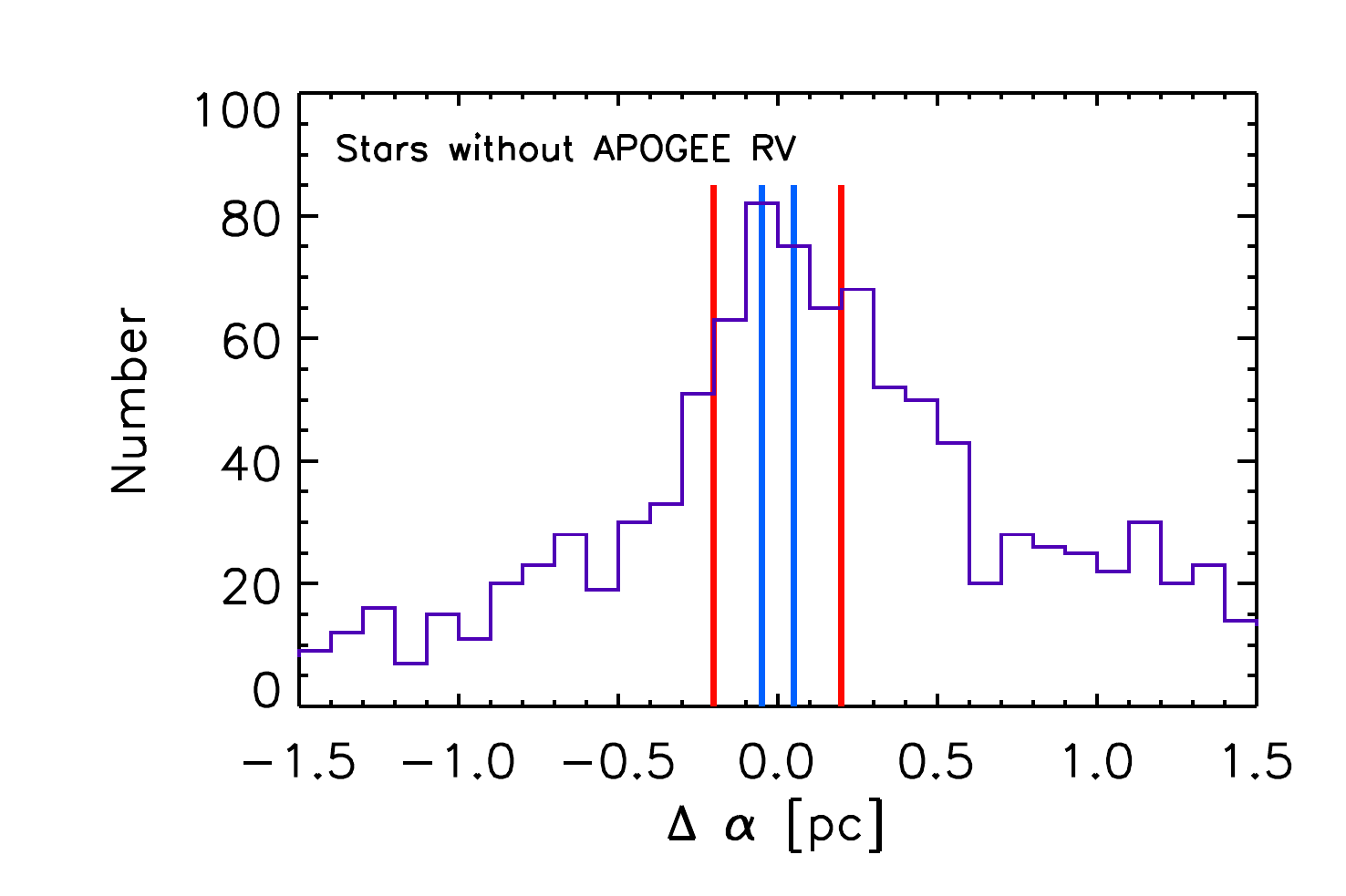}}\scalebox{0.70}{\includegraphics[trim = 22mm 0mm 0mm 0mm, clip]{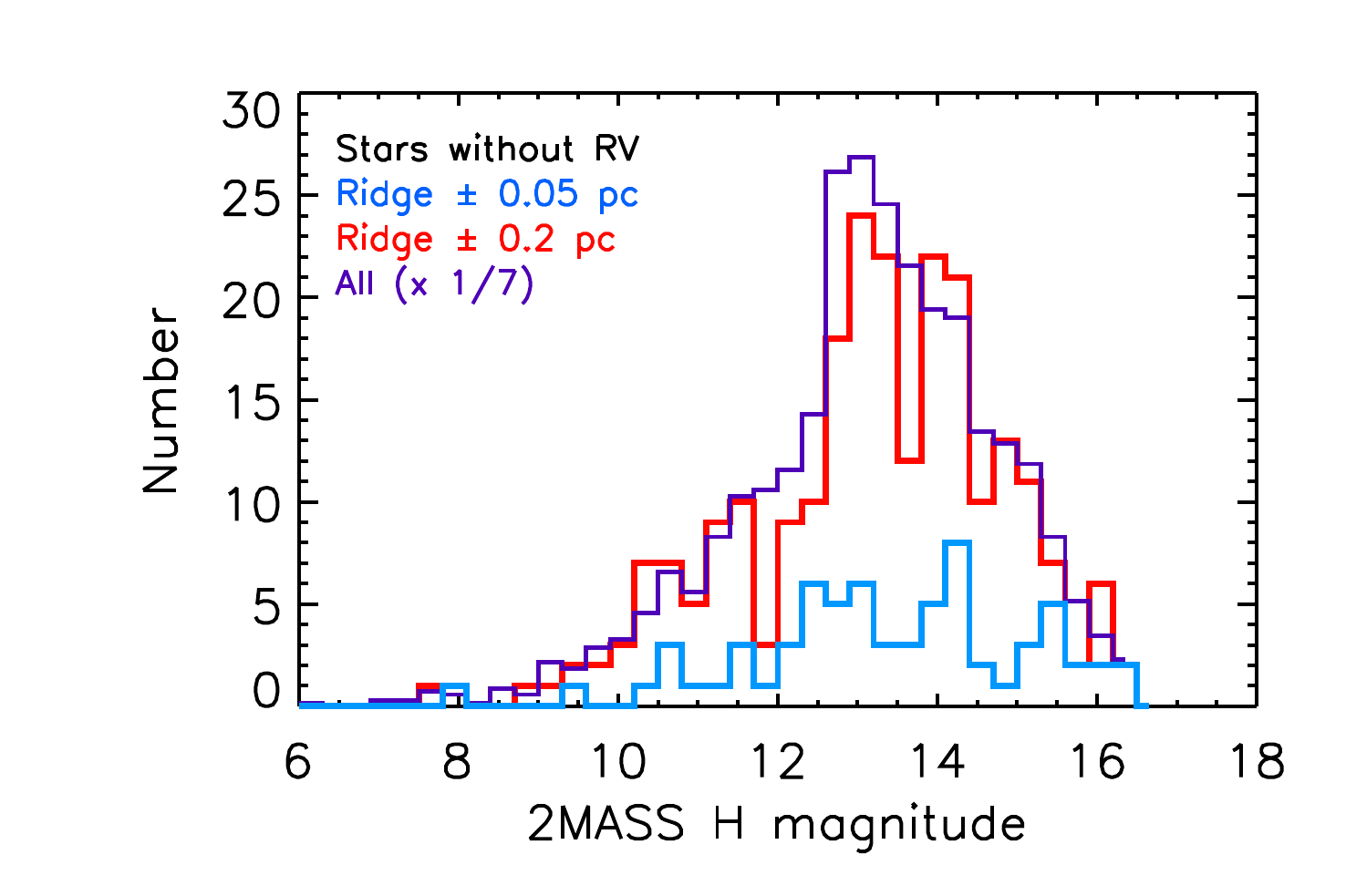}}
\caption{Left: Histogram of \citet{megeath12} IR-excess (Class II)
  stars that lack Apogee RVs as a function of distance from the
  filamentary spine.  Subsets within 0.2 and 0.05 pc are indicated.
  Right: $H$-band magnitude distribution of these three populations,
  with the whole population scaled down by a factor 7 for ease of
  comparison.  Full population and 0.2 pc subsample are statistically
  indistinguishable, nor is there any obvious bias from the $H<12.5$
  Apogee selection in the 0.05 pc subsample.  }
\label{fig:non-apo-hist}
\end{figure*}

We therefore further investigate this possible selection bias in
Figure~\ref{fig:non-apo-hist}, where we compare the $H$-band magnitude
distribution of all \citet{megeath12} IR-excess (Class II) stars that
lack Apogee RVs, with two subsets that lie projected close to the
filament ridgeline, within, respectively 0.2 and 0.05 pc.  The 0.2 pc
subsample looks identical to the (scaled) full sample, except that it
is noisier.  Hence, if there is any filament-extinction-induced bias
in the Apogee sample, it must be weak.  Nevertheless, given the
importance of the conclusions that we derive from the near (or
complete) absence of stars within the filamentary spines, we believe
that it would be valuable to obtain RVs at least for the 0.05 pc
sample. Note that reliable Apogee spectra can be obtained at
substantially lower signal-to-noise than the nominal threshold
\citep{ness15}, and this applies even more so if the main goal is to
obtain RVs.  

{\section{Characteristic Speeds and Times in Orion A}
\label{sec:charspeed}}

Figure~\ref{fig:dradrv} shows RV vs.\ $\alpha$ offsets from the
velocity and position ridgelines for three populations ISF protostars,
ISF stars, and L1641 stars.  It was introduced in
Section~\ref{sec:kinematics} in order to help assess whether stars
lying projected close to filamentary spines are actually in them.
Here we consider other implications of this figure.

The first point is that the shape of these distributions are,
respectively, highly elongated in the RV directions (ISF protostars),
moderately elongated (ISF stars), and slightly elongated (L1641
stars).  The axis scales are in ratios of $\pc/(\kms)\simeq 1\,$Myr in
all three cases.  This implies characteristic timescales of,
respectively, $\ll 1\,$Myr, $< 1\,$Myr, and $\lesssim 1\,$Myr.  These
timescales indicate either ages or turnaround times.

One way to distinguish between these possibilities is to examine the
phase-space distribution in these diagrams, which are color-coded by
$\delta$ in order to evaluate clumpiness in all three phase-space
coordinates.  If the stars or protostars had typically orbited once or
more (thereby impressing a turnaround timescale on the diagrams) then
they should be well mixed in the RV vs.\ $\Delta\alpha$ plots along
lines passing through the origin.  This is clearly not the case, with
lumpiness being most prominent in the protostar panel, but quite
prominent in the L1641 panel as well, which in particular has a severe
velocity asymmetry.

\begin{figure*}
\scalebox{0.57}{\includegraphics[trim = 10mm 0mm 0mm 0mm, clip]{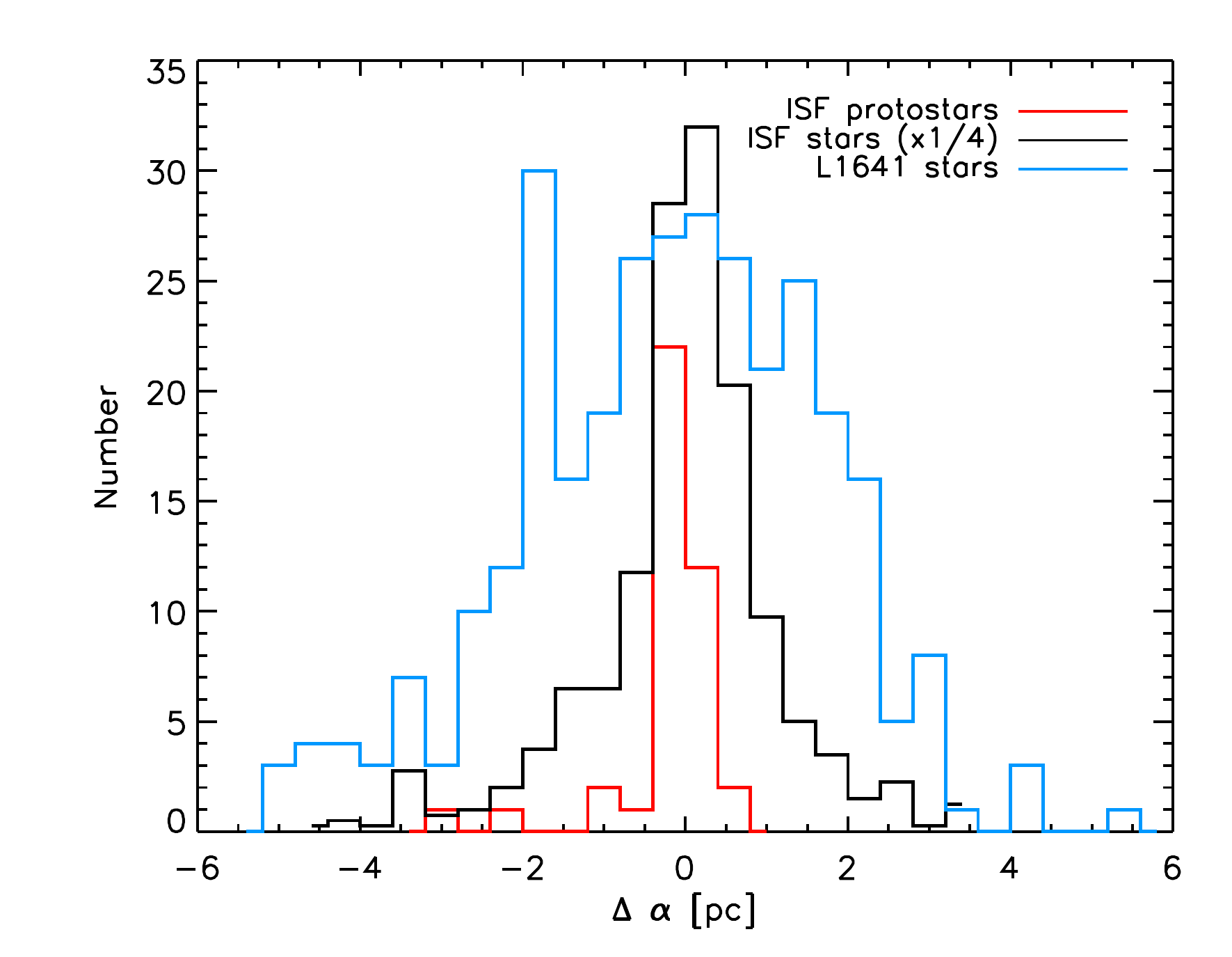}}\scalebox{0.57}{\includegraphics[trim = 20mm 0mm 0mm 0mm, clip]{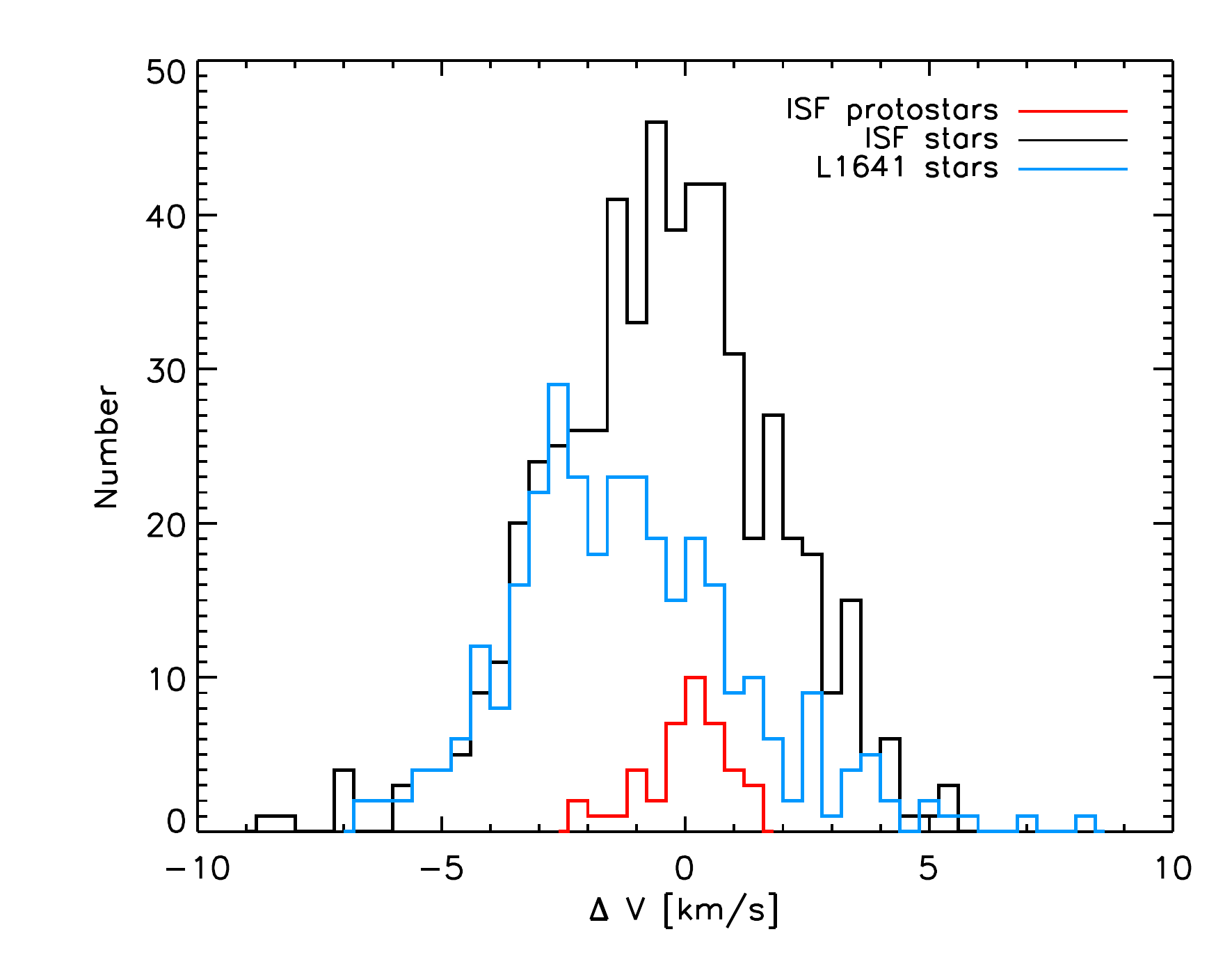}}
\caption{Histograms of offsets from spatial (left) and velocity
  (right) ridgelines for three populations ISF protostars (red), ISF
  stars (cyan) and L1641 stars (black).  For the two stellar
  populations, the ratios of the mean amplitudes of deviation given by
  Equation~(\ref{eqn:meanproj}) yield age estimates.  }
\label{fig:dradrvhist}
\end{figure*}

Another way to approach this issue is through the relations
derived in Section~\ref{sec:grav}.  Recall that for well mixed
orbits,  $\langle r \rangle \simeq r_\mx/\sqrt{2}$ and 
$\langle v\rangle=1.2\,\kms(r_\mx/\pc)^{3/16}$.  For quantities
observed in projection (i.e., those shown in Figure~\ref{fig:dradrv})
we expect both quantities to be reduced by $\sqrt{2}$.
Figure~\ref{fig:dradrvhist} shows histograms of these offsets
for all three populations.  Numerically we find mean projected
quantities
\begin{equation}
\langle r_\perp\rangle = (0.38,0.81,1.48)\,\pc;
\quad
\langle v_\parallel\rangle = (0.64,1.81,2.21)\,\kms
\label{eqn:meanproj}
\end{equation}
for the ISF protostars, ISF stars, and L1641 stars, respectively.
Here we have subscripted the two variables to emphasize that they
refer to different Cartesian directions. 

If, for example, the ISF stars were a well-mixed population, then
typically $r_\mx\sim 2*0.81\,\pc = 1.6\,\pc$.  Hence one would predict
$\langle v_\parallel\rangle\sim 1.2/\sqrt{2}(1.6)^{3/16}=0.9\,\kms$,
which is half the observed value.  For the L1641 stars, the
corresponding numbers are $r_\mx=3.0\,\pc$ and $\langle
v_\parallel\rangle = 1.05\,\pc$, i.e., less than half the observed
value.  From this we can conclude that both groups of stars have
typically not yet reached their first apogee after leaving the spine
of the filament.  We estimate mean ages (since ejection and consequent
termination of protostellar Class I phase) of these populations as
$0.81\,\pc/(1.81\,\kms)=0.44\,$Myr and
$1.48\,\pc/(2.21\,\kms)=0.65\,$Myr.

We note that these mean age estimates could be corrupted in various
ways.  For example, sources in binaries would have a random
component added to their motion.  This cannot be a dominant
effect since it would obliterate the clumpiness of the phase-space
diagrams if it were.  However, it could artificially depress the
age estimate.  The argument also assumes that ejections from
the filament are occurring at random angles relative to the filamentary
axis, which might not be the case.  In addition, the filament
has accelerated by different amounts in different directions
at each position along the filament since ejection.  This
adds noise (hence power) to both components.  This effect should
roughly cancel at first order but may still impact the estimates.

The same method cannot be applied to the protostars partly because
they lie too close to the filament in both position and velocity to
allow precise measurements of their offsets and partly because the
handful of protostars that form on sub-filaments are not being
properly handled by the formalism.  Indeed, the protostar panel shows
intriguing evidence for extreme hugging of the filament.  In
particular, there is a strong knot of protostars within $\pm
0.3\,\kms$ and $\pm 0.2\,\pc$ of the origin.  Since Class~I protostars
have estimated ages that are $\sim$\,3 times loger than Class~0
protostars \citep[e.g.,][]{dunham14} then we would expect to see
substantial diferences between the two sample if we where observing a
pure age effect.  However, although the numbers are small, we find
that the Class~0 and Class~I protostars exhibit no clear diferences in
projected quantities.

\section{Gravitational versus Magnetic Energy}
\label{sec:gravb}

Normally, one calculates the gravitational potential energy of
a system by asking how much energy must be injected to
separate its constituent particles to infinity.  We will
eventually have to make such an estimate, but of more immediate
relevance to understanding the waves that are propagating in the
filament is the energy required to lift all the material within
some radius $r_\mx$ that characterizes the filament oscillations
to that radius.  Recalling that $\rho(r)=(\gamma f/2\pi)K/pc^2(r/\pc)^{\gamma-2}$,
and $\Lambda(r) = fK(r/\pc)^\gamma$, we obtain
\begin{equation}
{dE\over dl} = \int_0^{r_\mx} dr\,(2\pi r\rho)2G\Lambda\ln{r_\mx\over r}
= {f^2\over \gamma}GK^2\biggl({r_\mx\over\pc}\biggr)^{2\gamma}
\label{eqn:dedl}
\end{equation}
We can now ask what is the rms magnetic field $B$ in this zone
that is required to equal this potential energy per unit length,
i.e., $\langle B^2\rangle \pi r_\mx^2/8\pi$.  We find
\begin{equation}
\sqrt{\langle B^2\rangle} = \sqrt{8G\over\gamma}f{K\over\pc}
\biggl({r_\mx\over\pc}\biggr)^{\gamma/2-1}
\rightarrow 70\,\mu G\biggl({r_\mx\over\pc}\biggr)^{-13/16}
\label{eqn:bg}
\end{equation}
This value is consistent with measurements by \citet{heiles97} at of
order $1\,\pc$ from the filament.  Equality of magnetic field and
gravitational potential energy would imply zero total energy on these
scales, even without considering kinetic energy due to turbulence, and
therefore would virtually guarantee instabilities.  Such instabilities
could plausibly trigger and sustain the transverse waves through the
filament that are indicated by its morphology, its kinematics, and the
quasi-periodic series of star-formation bursts in the north.

In general, one might be concerned that subcritical fields would
blow the system apart, perhaps allowing one burst of star formation
but not repeated ones.  However, the magnetic fields are only
subcritical on scales of the observed undulations.  As we have
shown in Section~\ref{sec:power}, the filament extends to much
larger radii than its easily visible spine.  Although we do not
know the full extent, we can place a lower limit on the ratio of
total gravitational binding energy to the local one by evaluating
Equation~(\ref{eqn:dedl}) at 8.5 pc and 1 pc
\begin{equation}
{(dE/dl)_{8.5}\over (dE/dl)_1}= 8.5^{2\gamma} = 5.0,
\label{eqn:dedl2}
\end{equation}
where $l$ is length along the filament.
Hence, it is quite possible for the magnetic fields to be subcritical
on the scales of the filament oscillations and supercritical on scales
of the filament as a whole.

\section{Discussion}

Starting from a new measurement (the first ever) of the gravitational
potential of Orion~A, and making use of information about the
kinematics of the gas, protostars, and stars, as well as the magnetic
fields, available from the literature, we have argued for a new
mechanism of star formation in the integral shaped filament (ISF),
namely, the slingshot mechanism.  

The slingshot arises specifically because the magnetic fields in this
region are subcritical on transverse scales of $\sim 1\,$pc,
leading to instabilities of two types.  First, on the (longitudinal)
scale of the filament as a whole ($\sim 10\,$pc), there are global
instabilities that result in a transverse wave that is apparent in the
positional and RV data.  Second, there are local pinching
instabilities, that generate repeated episodes of rapid star
formation, one of which is ongoing in the ONC, with previous episodes
having left behind the NGC 1977 and NGC 1981 clusters.  Protostars
form along the filament due to the high density of gas on its
ridgeline and are initially accelerated transversely with the filament
as it undergoes transverse oscillations.  This continues so long as
the protostars remain coupled to the gas.  Eventually, the mass of the
protostars becomes too great and they decouple.  At this point, they
continue to move with whatever transverse velocity the filament had at
the time of decoupling, while the filament itself continues to
oscillate.

The global instability (leading to the transverse wave) is long lived
because the magnetic fields are supercritical on scales of the
potential ($>8\,$pc transverse).  Hence, it survives many episodes of
star formation (as traced by the clusters we can see - NGC 1977 and
1981 -- and probably older ones that have subsequently dispersed).
The local instabilities are ``terminal'': they result in the
consumption and/or dispersal of the gas that is pinched off and are
responsible for the gradually growing gap between Orion~A and Orion~B
to the north.

Further south is L1641, which shares the same large-scale potential as
the ISF (see Figure~\ref{fig:cum2}), but has a significantly shallower
profile at transverse separations $b<2.5\,$pc.  In contrast to the
ISF, it does not exhibit transverse-wave morphology and it is not
forming stars at a rapid rate \citep{stutz15}.  In these respects, it
is more like Taurus and other nearby, lower-mass, molecular clouds.

That is, in our view, L1641 represents a ``first stage'' of star
formation, characterized by straight, magnetically supercritical
filaments with low star formation rates, whereas the ISF represents a
second stage, characterized by higher star formation rates and driven by
magnetic instabilities that give rise to transverse waves.  Thus, just
as the ISF's future is foretold by NGC 1977 and 1981, so the future of
L1641 is to collapse into a magnetically dominated filament with high
star formation rate.

However, in contrast to L1641, not all first-stage filaments will
reach the second stage.  L1641 can do so only because of its deep
potential well.  In this picture, lower-mass clouds like Taurus will
disperse without reaching the second stage.

Although this picture is completely new, it is largely consistent with
previous work on molecular clouds, both theoretical and observational,
as we now review.

\subsection{Two Mass Scales for Molecular Clouds}

To compare our measurements of the mass profiles of the ISF and L1641
with those of more nearby filaments that have been reported in the
literature, we focus on the projected mass per unit length within
$b<0.1\,$pc.  From Equation~\ref{eqn:lambdaA}, we find
$\lambda(0.1\,{\rm pc})=160\,M_\odot/$pc.  Because L1641 has a similar
profile for $b>2.5\,$pc but a shallower power law ($\gamma = 1/2$ vs.\
$\gamma=3/8$ for $b<2.5\,$pc) we estimate that the L1641 line density
is lower by a factor $(2.5/0.1)^{1/2-3/8}=1.5$.  That is,
$\lambda(0.1\,{\rm pc})=107\,M_\odot/$pc.

For Taurus, we derive $\lambda(0.1\,{\rm pc})=50\,M_\odot/$~pc from
Figure~5 of \citet{palmeirim13}.  For Musca, we derive
$\lambda(0.1\,{\rm pc})=25 \,M_\odot/$pc from Figure~5 of
\citet{kainulainen15}.  That is, the ISF has a factor 3--6 higher line
density that these nearby filaments.  It can be further contrasted in
that it has a much higher star formation rate than Taurus (and, of
course, than Musca, which is not forming stars at all).  Most
strikingly, Taurus and Musca are straight, whereas the ISF is integral
shaped.  Moreover, we detect no turnover in the profile at our
resolution limit of FWHM$\sim 0.04\,$pc, whereas more nearby clouds,
including Taurus and Musca, typically show FWHM$\sim 0.1\,$pc
\citep[see also e.g.,][]{andre14,arzoumanian11,konyves15}.  Hence,
this is another potential difference, although this issue must be more
carefully explored.  Finally, it appears that the density profiles of
more nearby clouds tend to cluster near $r^{-2}$ \citep[also
see][]{alves98,lada99} compared to $r^{-13/8}$ for the ISF.  However,
due to the somewhat irregular behavior of the nearby clouds as well as
relatively large errors in their measurement, this comparison cannot
yet be made rigorously.

Hence, with respect to all quantities and features that can be
reliably measured, the ISF is distinct from other nearby clouds.  This
is consistent with (although it by no means proves) our suggestion
that the ISF represents a different phase of star formation.  In this
picture, L1641 is intermediate, being governed by the same large-scale
potential, but not having (yet) evolved to the second phase
represented by the ISF.

To find analogs of the ISF, we must search to greater distances.  For
example, G11 has a line density of $\lambda = 600\,M_\odot$/pc out to
a column density of $\Sigma\geq 20\,M_\odot\rm /pc^2$
\citep{kainulainen13}.  To compare to the ISF, we consider an impact
parameter of $b=7.7\,$pc, which matches the same $\Sigma$ limit.  We
then find $\lambda(7.7\,{\rm pc})=830\,M_\odot/$pc for the ISF.  This
is qualitatively similar to G11.  Hence, while the ISF probably has
many analogs, essentially all of them are far more difficult to study.

\subsection{NGC 1333 in Perseus: An ISF Analog?}

\begin{figure}
\scalebox{0.4}{\includegraphics[trim = 0mm 0mm 0mm 0mm, clip, angle=0]{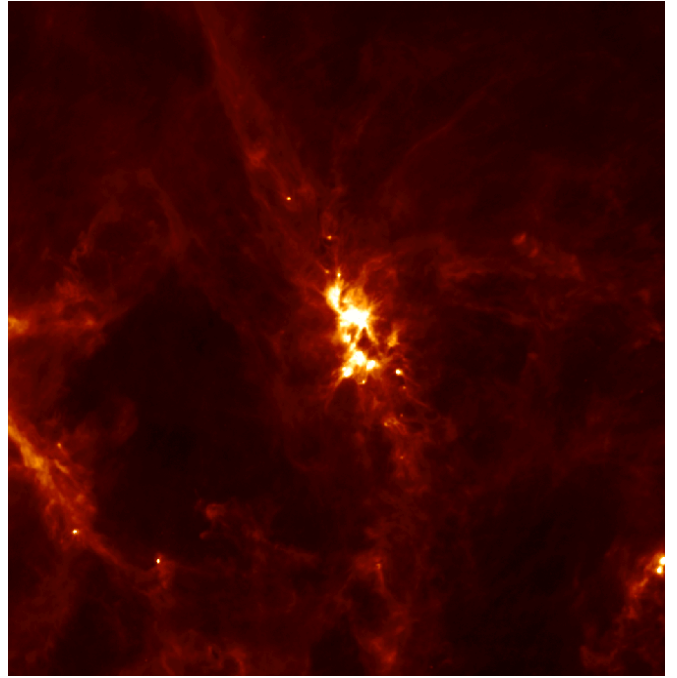}}
\begin{center}
  \caption{\herschel 250~$\mu$m image of the NGC 1333 region of
    Perseus, the scale is 6.1 pc on a side.}
\label{fig:ngc1333}
\end{center}
\end{figure}

Data and analysis by \citet{foster15} shows that NGC 1333 in Perseus
has a number of features that are remarkably similar to the ISF.
First, their Figure~5 shows RV oscillations, such as would be expected
from a transverse wave.  Second, the ``dense cores'' \citep{kirk07}
in this figure tightly follow this gas profile (just as the protostars
do in the ISF), while the stars show greater dispersion (just as the
ISF stars).  In addition, the line density in the filament may be
comparable to the ISF.  Like the ONC/ISF system, NGC 1333 is a dense
cluster embedded in a linear filament running north-south (see
Figure~\ref{fig:ngc1333}). However, in contrast to the ISF, there is
an important secondary east-west filament intersecting the primary
filament.  Perhaps for this reason, \citet{foster15} have analyzed the
profile using radial averages, so a precise quantitative comparison of
NGC 1333 and the ISF is not yet possible.

\citet{foster15} posit that the NGC 1333 dense cores may be coupled
to the magnetic fields, but do not suggest (as we have for the ISF)
that they inherit the magnetically accelerated gas motions when they
are released.  Instead, they argue that when the cores evolve into
stars, they directly acquire velocities characteristic of the
potential by some unspecified process.

\citet{foster15} have combined {\it Herschel} 160, 250, 350, 500
$\mu$m data to create a dust-column map of Perseus, similar to the one
made by Stutz et al.\ (2015) for Orion A, which was used in this
paper.  It would be straightforward to reconstruct the potential using
such a map.  We note that CephOB3, with a gas morphology dominated by
a main filament and clusters forming at each edge of the break in the
filament, may be a promising ISF analog.  See Figure~2 of
\citet{gutermuth11}.

\subsection{Molecular Cloud Formation}

Low mass molecular clouds are characterized by straight, sometimes
intersecting filaments.  For example, \citet{palmeirim13} observe a
large straight filament in Taurus (B211/3), with a radial density
profile $\rho \propto r^{-2\pm 0.4}$ between $r = 0.05$~pc to 4~pc and
mass per unit length $\lesssim 45 M_\odot$/pc.  The e.g.\ Aquila,
Lupus, and IC5146 filaments
\citep{konyves15,benedettini15,arzoumanian11} also appear fairly
straight.

Such structures are readily produced in simulations of turbulent gas
with and without magnetic fields
\citep{padoan95,maclow04,federrath15,smith14,kirk15,chen15,smith16}.
The fact that plausible representations of filaments can be produced
even without magnetic fileds makes turbulence a leading candidate for
the formation of molecular clouds.

Of course, the situation cannot be so simple because magnetic fields
are observed to be almost universally associated with filaments.  For
example, \citet{palmeirim13} observed striations perpendicular to the
Taurus B211/3 filament and argued that these were tracing mass
accretion.  These striations are approximately parallel to the
projected plane of the sky magnetic field, as traced by optical
polarization
\citep{chapman11,heiles00,heyer08}. \citet{fiege00,fiege00b} argue
that such a ``shallow'' $(r^{-2})$ density profile as that estimated
in B211/3 would be consistent with models of filaments in isothermal
equilibrium that are threaded by helical magnetic fields.

However, the ubiquity of magnetic fields is not necessarily
inconsistent with the magnetically-free turbulent simulations.  It
could be that when filaments form out of the ISM, they contain
magnetic fields but these are subdominant.  Hence, simulations need
not take account of them to get the basic structure right.  Then
later, as gas accretion concentrates the field lines, the magnetic
fields play an increasingly central role.

This would be consistent with our picture, in which magnetic fields
play a qualitatively greater role in the second stage of star
formation (illustrated by the ISF) than in the first stage, as
exemplified by L1641 \citep[e.g.,][]{polychroni13} and also the
lower-mass, nearby clouds.  These, indeed, all have nearly
straight-line filaments.  While they may have magnetically-mediated
flows, none show evidence for the violent magnetic instabilities that
characterize the ISF.

And, in conformity with this, no turbulence simulation has ever
produced a filament that appears ``integral shaped''.  Moreover, it is
a definite prediction from our picture that none ever will, whether
they include magnetic fields or not.  We argue that the ISF is a
specific product of internal evolution of the cloud, and cannot be
directly produced by any combination of initial conditions, with or
without magnetic fields.

\subsection{Molecular Cloud Lifetimes}

\citet{hartmann01} and \citet{ballesteros07} argue that lifetimes of
nearby low mass molecular clouds are $\sim$~1 to 3~Myr.  These
lifetime estimates hinge on two facts.  First, the stars forming in
these clouds have ages of order $\sim$~1 to 3~Myr and not more
\citep{jeffries11}.  Second, of all clouds having qualitatively
similar surface densities and masses, roughly two-thirds are forming
stars.  Hence, clouds must start forming stars soon after they
coalesce and they must disperse soon after they start forming stars.

In seeming contrast, \citet{murray11} (see also Meidt et al.\ 2015)
measures lifetimes 27$\pm$12~Myr for massive molecular
clouds. However, he argues that there is no real conflict.  Partly he
takes issue with stellar-based age estimates of the lower mass clouds,
but his main point is that the cloud masses are an order of magnitude
larger, while their densities are a factor few lower, leading to
longer infall times and so lifetimes.

Within the framework of our picture, these results are consistent for
a fundamentally distinct, though related, reason.  The low-mass clouds
that were the basis of the \citet{hartmann01} study have insufficient
mass to form (or to survive the formation of) magnetically dominated
filaments.  Hence, they are dispersed as soon as they start forming
stars.  The massive clouds studied by \citet{murray11} do form such
structures.  As a result, they are capable of the higher rates of star
formation, such as that seen in the ISF, and this is the reason that
they are noticed despite their typical distances of several kpc.
However, because they are so far away, they cannot be studied in the
same detail as Orion A, and hence the fact that they are in a second
phase of star formation has never been recognized.

\subsection{Violent Acceleration: Magnetic Fields vs.\ Cold Collapse}

We have shown that the ISF stars are moving substantially more rapidly
and display much less clumping than ISF protostars.  \citet{foster15}
report similar results for NGC 1333 in Perseus.  Somehow, stars are
being violently accelerated post birth.  The question is how?

To date, there are only two theories that purport to explain this.  We
have proposed a slingshot mechanism by which magnetically driven
instabilities lead to ejection of protostars from an oscillating
filament.  \citet{hartmann07} have proposed violent relaxation
following cold collapse of a star-forming cloud of gas \citep[see
also][]{tobin09}.

How do these theories relate to the facts?  First, all of the
available evidence is consistent with filaments being wrapped in
toroidal magnetic fields.  In all star-forming filaments that have
been studied, the transverse fields (traced by polarization) are
roughly perpendicular to the filament, which is exactly what is
expected from toroidal fields.  For the ISF, the line of sight fields
change direction as one crosses the filament, which is also the
expected behavior.  The energy of the magnetic field is comparable
to the gravitational potential and so is high enough to generate
instabilities.  The filament's appearance is consistent with it
undergoing a transverse wave.  Such an accelerating filament will
naturally hold on to diffuse material that is in the process of
forming a protostar and will equally naturally release it once the
protostar becomes sufficiently compact.

In cold collapse, the protostars must also form on filaments.
Otherwise we live in a special epoch, not only for the ISF but also
NGC 1333.  These filaments then must have subsequently disappeared,
being replaced by the filament that we can currently see, which is one
of the most prolific in the nearby universe.  In addition, simulations
carried out by \citet{kuznetsova15} show that the ``sink particles''
(representing protostars) do not cluster in phase space like the real
ISF protostars (see their Figures 1, 2, and 6 and our
Figure~\ref{fig:morph-proto}).

One might counter that if there were strong toroidal fields, then
these would have prevented formation of the filament in the first
place by blocking infall.  Our answer: subcritical magnetic fields
are there.  \citet{heiles97} measured their amplitude 20 years ago and
we have now shown they are subcritical by measuring the potential.
Such high field strengths are naturally explained by magnetic
compression of gas due to currents.  That is, the material gets to its
current position not by crossing field lines but by compressing them.
It may be objected that this process would lead to pinching
instabilities.  Our answer: yes, pinching instabilities are expected
and this is exactly what leads to cluster formation.

\section{Conclusions}

Systematic review of publicly available data on Orion A has led to
striking conclusions.  Some of these follow directly from the
observations.  The surface density of the filament declines as a power
law as one moves away from its spine, which itself displays previously
well-known undulations.  The regularity of the larger structure
already tells us that the dense spine at its center is being subjected
to strong transverse perturbing forces, since its relaxation time is
much shorter than that of the larger, regular structure.  The
undulations are present in both the position and radial velocity of
the spine.

These undulations appear to be ejecting protostars, which is probably
the mechanism for shutting off their accretion.  The evidence for this
is that protostars, virtually without exception, all lie projected
directly on filaments and their radial velocities are consistent with
those of this filamentary gas at $<1\,\kms$, while for stars the
opposite is the case: they are generally both off filaments and have
radial velocities that differ from the neighboring filaments by
several $\kms$.  Of course, this argument can be directly applied only
to the subset with RV measurements, and this subset could in principle
have been biased against stars inside filaments due to higher
extinction.  However, our investigation showed that any such bias must
be very weak.

The relatively high ratio of velocity-to-physical offsets of ISF stars
from their respective ridgelines leads to a relatively low age
estimate $\sim 0.44\,$Myr, where ``age'' means time since ejection
from the filament and consequent termination of the Class I protostar
phase.  This is consistent with other evidence of their youth such as
the high ratio of protostars to stars and the functional form of the
column-density histogram \citep{stutz15}.

The youth of the ISF is of particular interest because it fits into a
larger pattern.  The ISF sits at the northern end of the Orion A
filament, where ``end'' means end of a discernible spine of gas.  Yet
just 1.6 pc from the ``northern tip'' of the ISF (4.1 pc from the ISF
center) is the young cluster NGC 1977, which we dub the ``orphan
cluster'' because the gas column from which it formed is no longer
present.  The cluster is predominantly composed of disk sources (Class
II stars), from which we estimate an age of 2 Myr.  Moreover, 2.9 pc
further north is an older cluster, which \citet{megeath13} showed is
predominantly composed of photospheric (Class III) stars, with a
minority of Class II stars.  Hence, it is yet older, perhaps 4 Myr.
Thus we see the presence and/or vestiges of three roughly periodic
episodes of star formation, each of which took place at the northern
end of the Orion A filament as it existed at the epoch of that
cluster's formation.  Although there is no direct evidence for
extrapolating this trend, one can guess that these quasi-periodic
episodes of cluster formation stretch back further in time and are
responsible for the progressive dispersal (and consumption) of the gas
filament that used to extend all the way to Orion B.  Indeed, we see
some evidence for dispersal of the gas associated with the ``orphan
cluster'' in the westward trail of gas that we have dubbed the ``ghost
filament''.

In order to assess the cause of these periodic bursts of cluster
formation, we construct the simplest possible model: we assume that
the observed surface density power law $\Sigma(b)\propto b^{-5/8}$ is
due to an underlying axisymmetric density profile
$\rho(r)\propto r^{-13/8}$.  It is then straightforward to determine
the integrated line density as a function of cylinder radius, the
potential energy, etc.  We find that the line density of gravitational
potential energy on 1 pc scales is comparable to the line density of
magnetic energy as measured by \citet{heiles97} on similar scales.
That is, the magnetic fields are subcritical, which naturally leads
to repeated perturbations propagating through the filament as mainly
traverse waves.  We also show that while the magnetic fields are
subcritical on the 1 pc scales of the observed undulations of the
filament, they are supercritical on the scales of the extended regular
structure, which we trace to at least 8.5 pc.  Thus, the spine can be
subjected to violent periodic perturbations, while at the same time
surviving many such episodes.

It is the transverse acceleration of the filament spine that is
responsible for ``ejecting'' the protostars via a ``slingshot''
mechanism.  That is, it is not the protostars that leave the gas but
the gas that leaves the protostars.  As long as the protostars are
entrained in extended envelopes that are mechanically linked to the
larger spinal column, the cores at their centers are dragged along
with the column as it accelerates.  However, when the cores reach
sufficient mass relative to the larger envelope, their own inertia
prevents them from following the gas as it accelerates.  See
Figure~\ref{fig:cartoon} for a cartoon illustration of this process.
If this phase occurs at a time of low acceleration, then even a
tenuous connection between the core and the envelope is sufficient to
keep the protostar in the column.  However, at times of high
acceleration, ejection can occur at an earlier phase of protostar
evolution.

We therefore predict that the initial launching velocities of the
protostars should be of order the semi-amplitude of the observed
undulations of the spine, namely $\sim 2.5\,\kms$.  Given the
potential $\Phi(r) = 6.3(\kms)^2(r/\pc)^{3/8}$, this implies maximum
excursions of $r_\mx\sim (2.5^2/2\times 6.3)^{8/3}\sim 4.0\,\pc$,
which are achieved after a time $\Delta t\sim 4.8\,$Myr.  Thus, most
recently formed stars should be on their first ascent away from their
birth filament, a prediction that could be confirmed by Gaia proper
motions.

The timescales of these filament motions can be estimated from the
ratio of amplitudes of spatial to velocity amplitudes, namely
$\sim 1.5\,\pc/(2.5\,\kms)\sim 0.6\,$Myr.  Hence, the accelerations
are $a\sim 4\,\kms\,{\rm Myr}^{-1}$.  This is similar to the
acceleration at $\sim 0.05\,\pc$ from a $\sim 2\,M_\odot$ protostar.
Since the filament ridges have similar widths to the first number and
protostar envelopes have similar masses to the second, it may be that
it is the amplitude of these undulations that sets the stellar mass
scale.

\begin{acknowledgements}

  We thank James Di Francesco for providing the NH$_3$ data in advance
  of publication.  We thank Boaz Katz, Ralf Pudritz, and Martin Pessah
  for excellent discussions that improved this work.  We thank the
  anonymous referee for a helpful report. We thank Melissa Ness, Sarah
  Sadavoy, Tom Megeath, Lee Hartmann, and Andrey Kravtsov for
  informative and helpful discussions.  AG is grateful for the
  hospitality of the MPIA and acknowledges support from NSF AST
  1516842.

This paper includes data from \herschel, a European Space Agency (ESA)
space observatory with science instruments provided by European--led
consortia and with important participation from NASA.  

We acknowledge the use Digitized Sky Survey data in this work. The
full DSS acknowledgement can be found at
https://archive.stsci.edu/dss/acknowledging.html

We acknowledge the use of SDSSIII data in this work.  The full SDSSIII
acknowledgement can be found at https://www.sdss3.org

We acknowledge the use of the Georgia Institute of Technology website
Edward Emerson Barnard's Photographic Atlas of Selected Regions of the
Milky Way.  The full acknowledgement can be found at
http://www.library.gatech.edu/barnard/credits.html

\end{acknowledgements}

\end{document}